\documentclass{article}
\usepackage{epsfig,amssymb,float,psfig}
\input{epsf}
\newcommand{\be}{\begin{equation}}
\newcommand{\ee}{\end{equation}}
\newcommand{\ba}{\begin{eqnarray}}
\newcommand{\ea}{\end{eqnarray}}
\newcommand{\beq}{\begin{equation}}
\newcommand{\eeq}{\end{equation}}
\newcommand{\beqa}{\begin{eqnarray}}
\newcommand{\eeqa}{\end{eqnarray}}

\newcommand{\vs}{\vspace{-0.15cm}}
\def\barre#1{{\not\mathrel #1}}

\begin{document}

\begin{flushright}
{\tiny  FZJ-IKP(TH)-2002-20} \\
{\tiny  $\,$} \\
\end{flushright}

\vspace{1cm}

\begin{center}

\bigskip

{{\Large\bf Spin structure of the nucleon at low energies
}}

\end{center}

\vspace{.3in}

\begin{center}
{\large
V\'eronique Bernard$^{\dagger,}$\footnote{email: bernard@lpt6.u-strasbg.fr},
Thomas R. Hemmert$^{\ast,}$\footnote{email: themmert@physik.tu-muenchen.de},
Ulf-G. Mei{\ss}ner$^{\ddagger,}$\footnote{email:
  u.meissner@fz-juelich.de} \footnote{Address after Jan. 1$^{\rm
    st}, 2003$: Helmholtz Institut f\"ur Strahlen- und Kernphysik (Theorie),
    Universit\"at Bonn, Nu{\ss}allee 14-16, D-53115 Bonn, Germany.}}

\vspace{1cm}

$^\dagger${\it Universit\'e Louis Pasteur, Laboratoire de Physique
               Th\'eorique\\
               F--67084 Strasbourg, France}

\bigskip

$^{\ast}${\it Technische Universit\"at M\"unchen, Physik Department T-39\\ 
              D-85747 Garching, Germany }

\bigskip

$^\ddagger${\it Forschungszentrum J\"ulich, Institut f\"ur Kernphysik
(Theorie)\\ D-52425 J\"ulich, Germany}

\bigskip

\end{center}

\vspace{.6in}

\thispagestyle{empty}

\begin{abstract}\noindent
The spin structure of the nucleon is analyzed in the framework of a 
Lorentz--invariant formulation of baryon chiral perturbation theory.
The structure functions of doubly virtual Compton scattering are
calculated to one--loop accuracy (fourth order in the chiral expansion).
We discuss the generalization of the Gerasimov--Drell--Hearn sum rule,
the Burkhardt--Cottingham sum rule and moments of these. We give predictions
for the forward and the longitudinal--transverse spin polarizabilities
of the proton and the neutron at zero and finite photon virtuality. 
A detailed comparison to results obtained in heavy baryon chiral 
perturbation theory is also given.
\end{abstract}

\vfill

\pagebreak

\section{Introduction and summary}
\def\theequation{\arabic{section}.\arabic{equation}}
\setcounter{equation}{0}
\label{sec:intro}

Understanding the spin structure of the nucleon is a central topic of
present nuclear and particle physics activities, for a review see
\cite{review}. Of particular interest are certain sum rules which
connect information at all energy scales, like e.g. the
Gerasimov--Drell--Hearn (GDH) sum rule and its
generalization to finite photon virtuality or the Burkhardt--Cottingham (BC) 
sum rule. Such sum rules are interesting from the theoretical point of view because 
they constitute moments of the sought after nucleon spin
structure functions $g_1$ and $g_2$. On the experimental side
challenging new meson production experiments using real or virtual photons play an
important role  since only
recently it has become possible to work with polarized beams and polarized targets,
thus offering the possibility of mapping out the nucleons' spin structure
encoded in these two functions, which can be
formulated on a purely partonic (high energy regime) or hadronic level
(low energy regime). In both these extreme cases, systematic and
controlled theoretical calculations can be performed. The region of
intermediate momentum transfer  is accessible using quark/resonance
models or can be investigated using dispersion relations. In fact, one
of the final goals of many of these investigations is to obtain an
understanding of how in  QCD this transition from the
non--perturbative to the perturbative regime takes place, guided by the precise
experimental mapping of spin--dependent observables from low momentum transfer to
the multi--GeV region, as it is one of the main thrusts of the research carried
out e.g. at Jefferson Laboratory. 

\medskip\noindent
Here we focus on a theoretical investigation of the nucleon's spin structure in the
non-perturbative regime of QCD.
We are utilizing chiral perturbation theory (CHPT) to analyze the structure
of the nucleon  at low energies, based on the spontaneous and explicit
chiral symmetry breaking QCD is supposed to undergo (for a general
review, see e.g. \cite{BKMrev}). By now it is well established that the
pion cloud plays an important role in understanding the nucleons properties
in the non--perturbative regime of QCD, and many processes have been analyzed
using chiral perturbation theory. Some recent work in various
versions/extensions of baryon CHPT pertinent
to the topics discussed here can be found e.g. in 
Refs.~\cite{BKMdhg,EKPW,Ji1,Ji2,VB,Kao}.
In the letter \cite{BHMdhg} we had presented a novel analysis of
chiral loop effects in the generalized Gerasimov--Drell--Hearn sum
rule in the framework of a Lorentz--invariant formulation of baryon
chiral perturbation theory \cite{BL}. We performed a complete one--loop
calculation for the spin--dependent proton and neutron structure 
functions $S_{1,2} (\nu =0 , Q^2)$ of forward double virtual Compton scattering 
(V$^2$CS),  where $\nu$ is the energy transfer
and $Q^2$ the negative of the photon virtuality (momentum transfer
squared). Combining the methods of CHPT and analyticity, one can deduce all desired
spin-related  sum rules, respectively all moments of the nucleon's 
spin structure functions, from these  two basic building blocks for 
$Q^2 \lesssim 0.3$~GeV$^2$.
Results where thus presented for the V$^2$CS structure functions
$\bar S_1^{p,n} (0,Q^2)$ (where the bar indicates the subtraction of the elastic
intermediate state)  and for the first moment of the sum rule for the
spin--dependent structure function $g_1$ (for precise definitions, see
below). Also,  a comparison to the previously obtained so called ``heavy baryon''
results, which are implicitly contained in our calculation and can be extracted by 
performing a 1/(nucleon mass) expansion of our relativistic results, was given. 
Here, in the light of the sometimes rather slowly converging $1/m$ expansion 
of the heavy baryon framework in the spin sector \cite{BHMdhg}, we do not only give 
a detailed exposition of how the relativistic 
results have been obtained, but also give novel results for the 
structure functions $\bar S_2^{p,n} (0,Q^2)$, for the generalization of the DHG and
BC sum rules  as well as for the forward and longitudinal--transverse
spin polarizabilities. These latter two quantities  parameterize the
forward spin--dependent (virtual) Compton scattering amplitude at low energies and
encode important information about higher moments of the nucleon's spin structure 
functions.

\begin{figure}[H]
\vspace{0.9cm}
\parbox{.49\textwidth}{\epsfig{file= 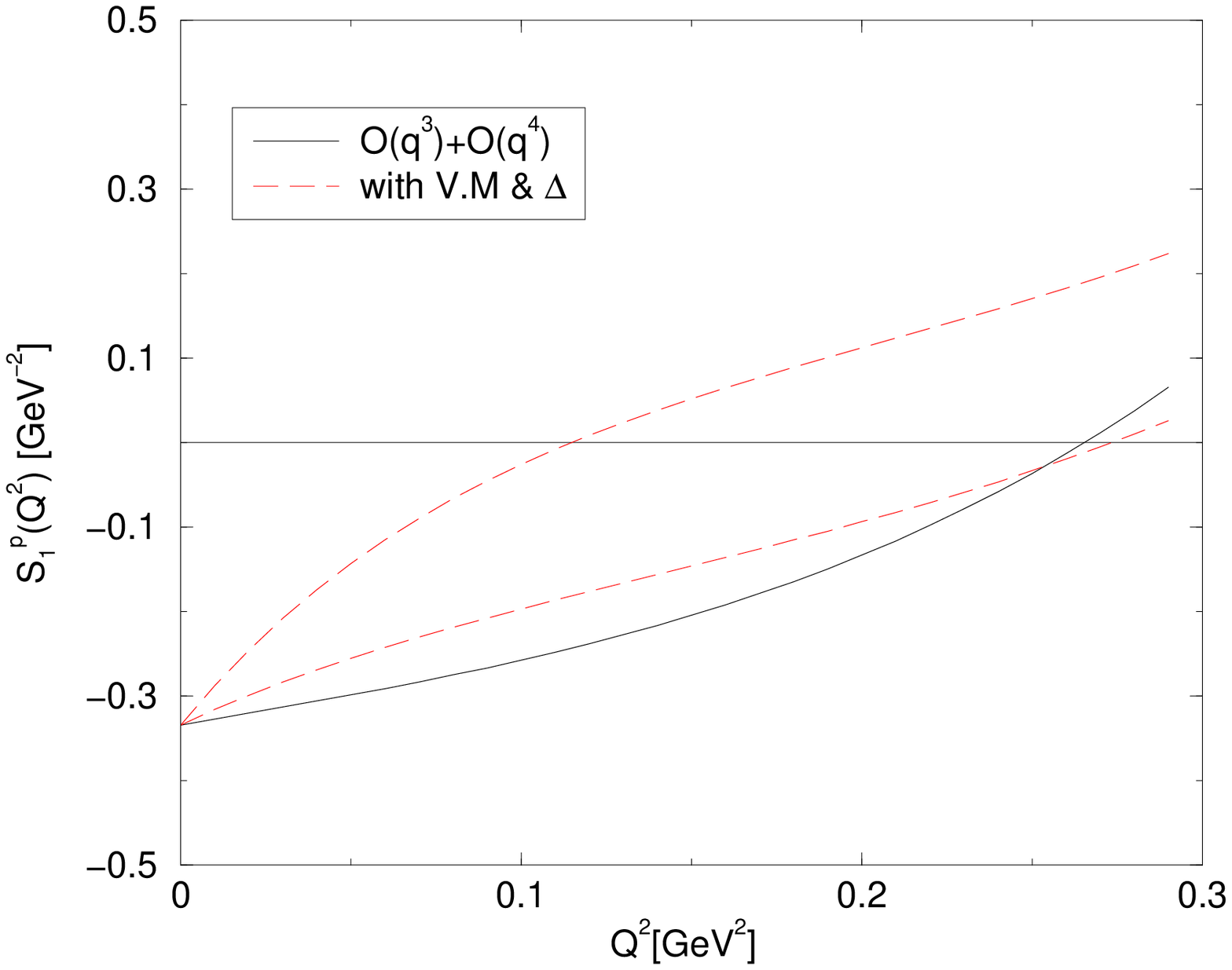,width=.48\textwidth,silent=,clip=}}
\hfill
\parbox{.49\textwidth}{\epsfig{file= 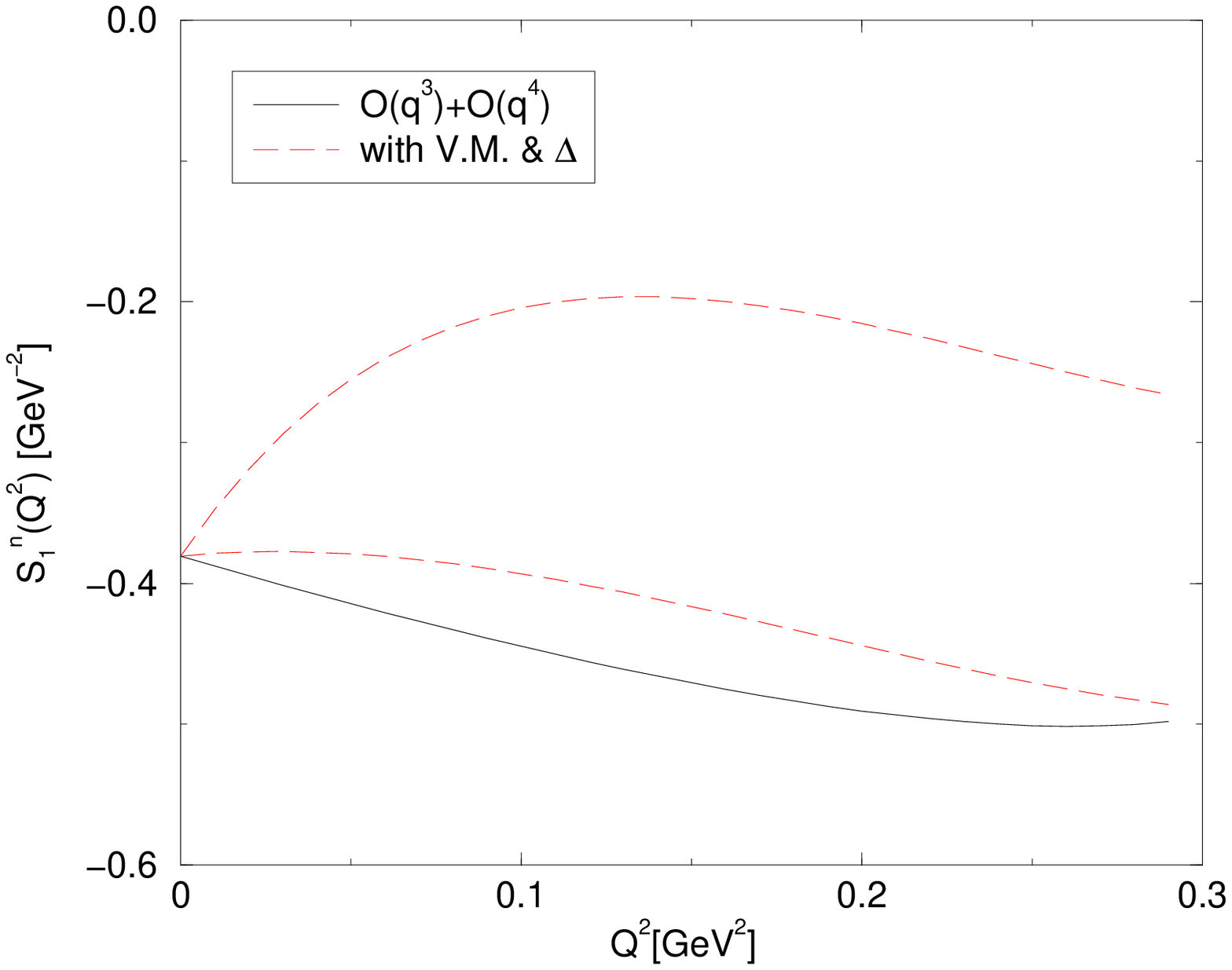,width=.48\textwidth,silent=,clip=}}
\vspace{0.3cm}
\begin{center}
\caption{Predictions for the structure function $\bar S_1 ({0,Q^2})$ with the elastic
intermediate state subtracted in units of GeV$^{-2}$. 
The solid lines show the fourth order (one--loop) result
whereas the bands given by the  dashed lines refer to the one--loop plus resonance
(delta Born graphs and vector meson) results. Left (right) panel: Proton (neutron).
\label{fig:s1pn}}
\end{center}
\end{figure}

\begin{figure}[H]
\vspace{0.9cm}
\centerline{
\epsfysize=2.72in
\epsffile{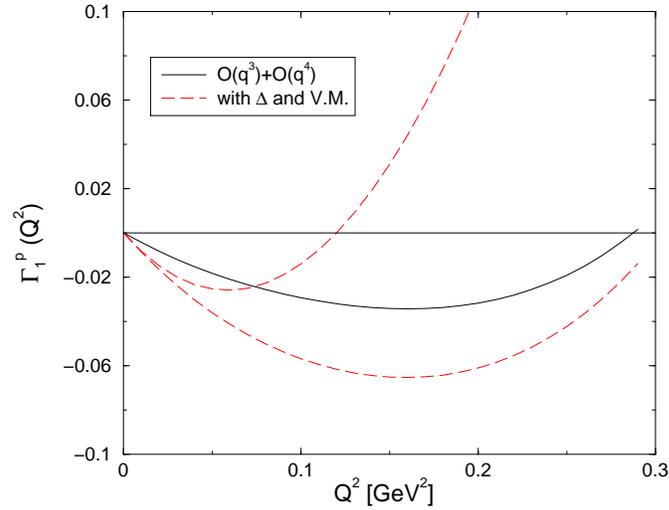}
}
\vspace{0.3cm}
\begin{center}
\caption{
Prediction for the moment $\Gamma_1^p ({Q^2})$.
The solid lines show the fourth order (one--loop) result
whereas the bands given by the  dashed lines refer to the one--loop plus resonance
(delta Born graphs and vector meson) results. No $N\Delta \gamma$ form factor
was used.
\label{fig:gamma1p}}
\end{center}
\end{figure}

\begin{figure}[H]
\vspace{.5cm}
\parbox{.49\textwidth}{\epsfig{file= 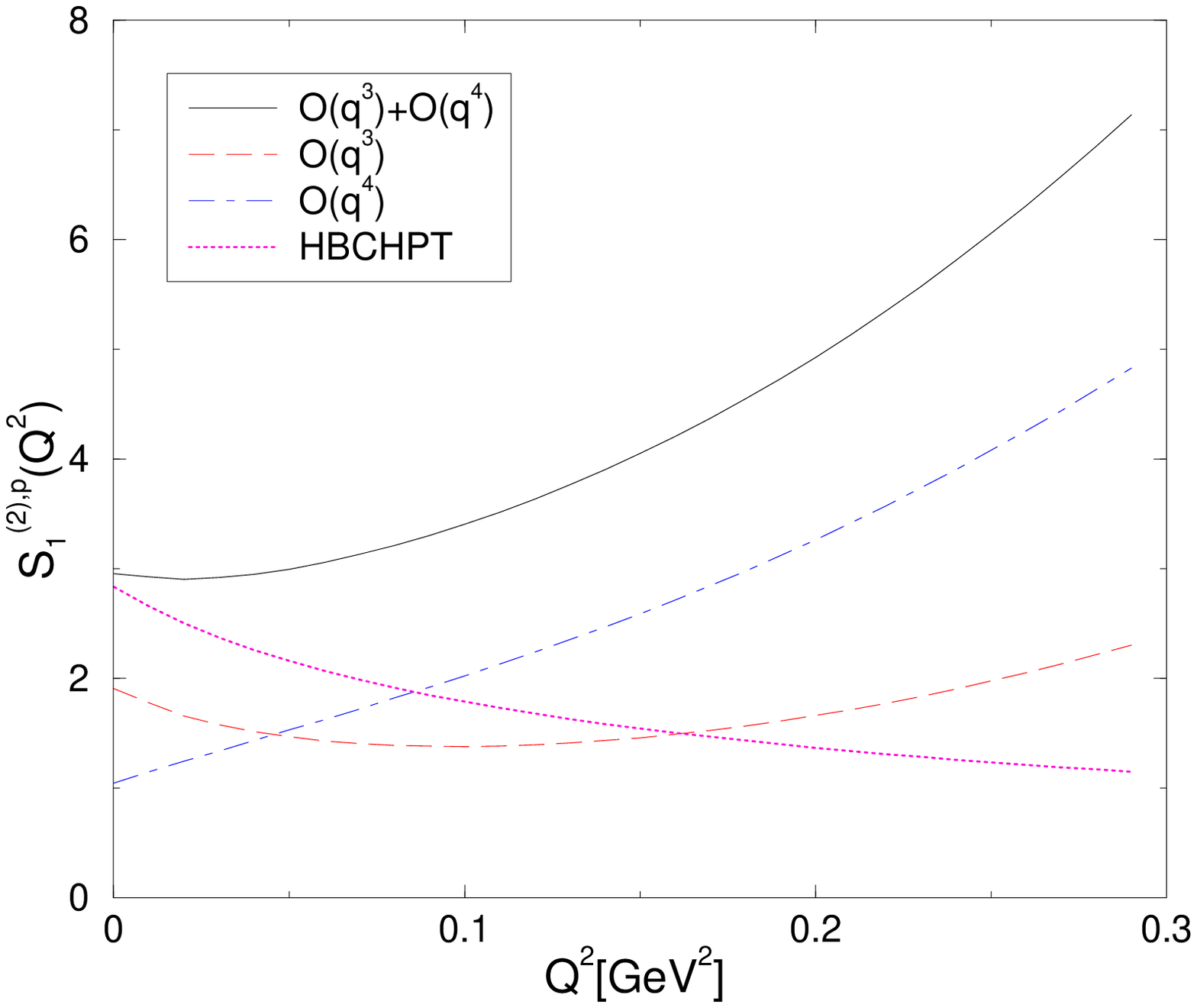,width=.44\textwidth,silent=,clip=}}
\hfill
\parbox{.49\textwidth}{\epsfig{file= 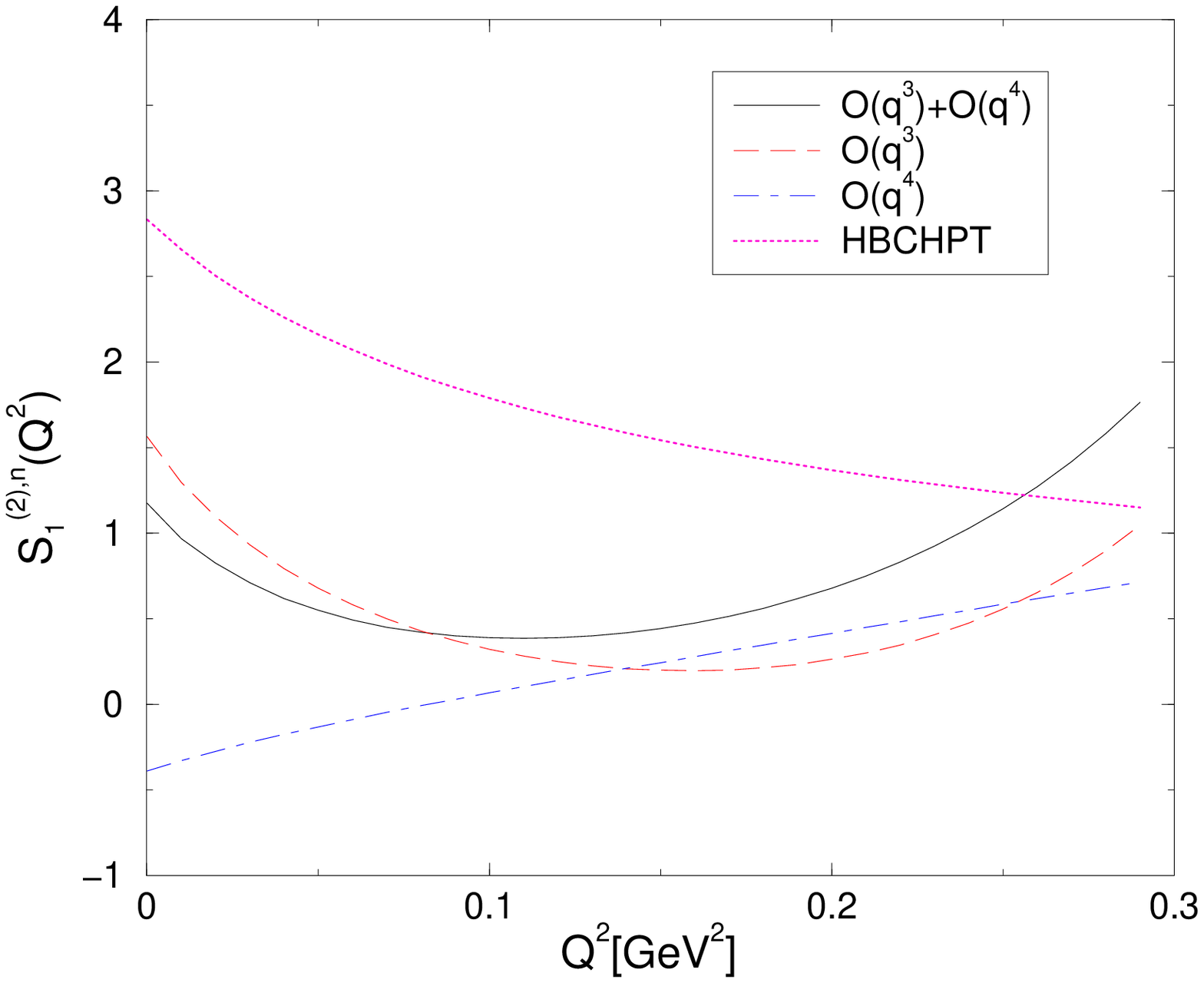,width=.44\textwidth,silent=,clip=}}
\vspace{0.3cm}
\begin{center}
\caption{\label{fig:s12pn}
          Chiral loop contribution to the first moment $\bar{S}_1^{(2)}
          (0,Q^2)$ with the elastic contribution subtracted
          in units of GeV$^{-4}$. 
          The solid (dashed/dot-dashed) line gives the result of
          the present calculation to order $q^4$ (the contribution
          from the third/forth order) in comparison
          to the heavy baryon result of \protect\cite{Ji2} (dotted line).
          Left (right) panel: Proton (neutron).
}
\end{center}
\end{figure}

\begin{figure}[H]
\vspace{.5cm}
\parbox{.49\textwidth}{\epsfig{file= 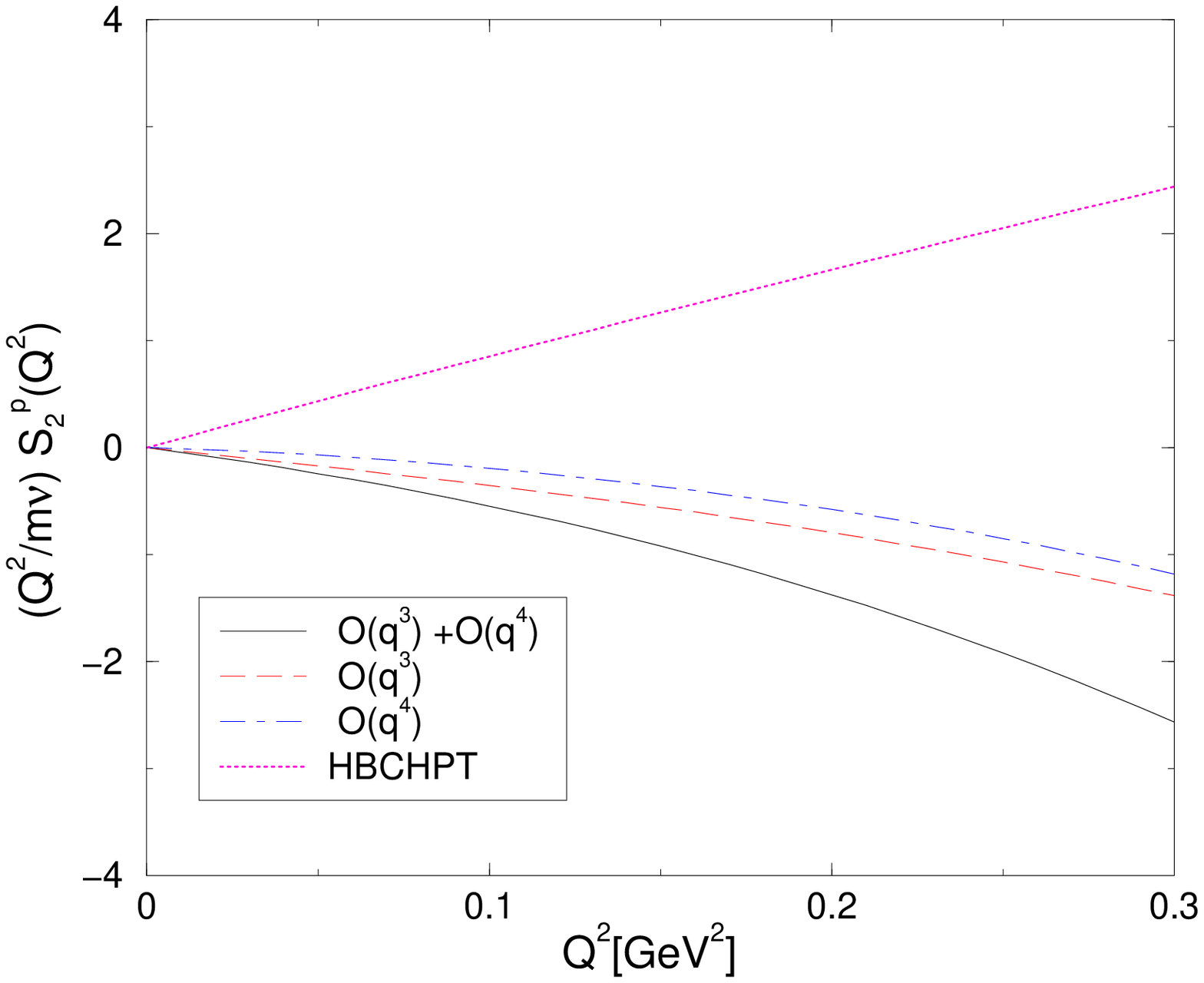,width=.45\textwidth,silent=,clip=}}
\hfill
\parbox{.49\textwidth}{\epsfig{file= 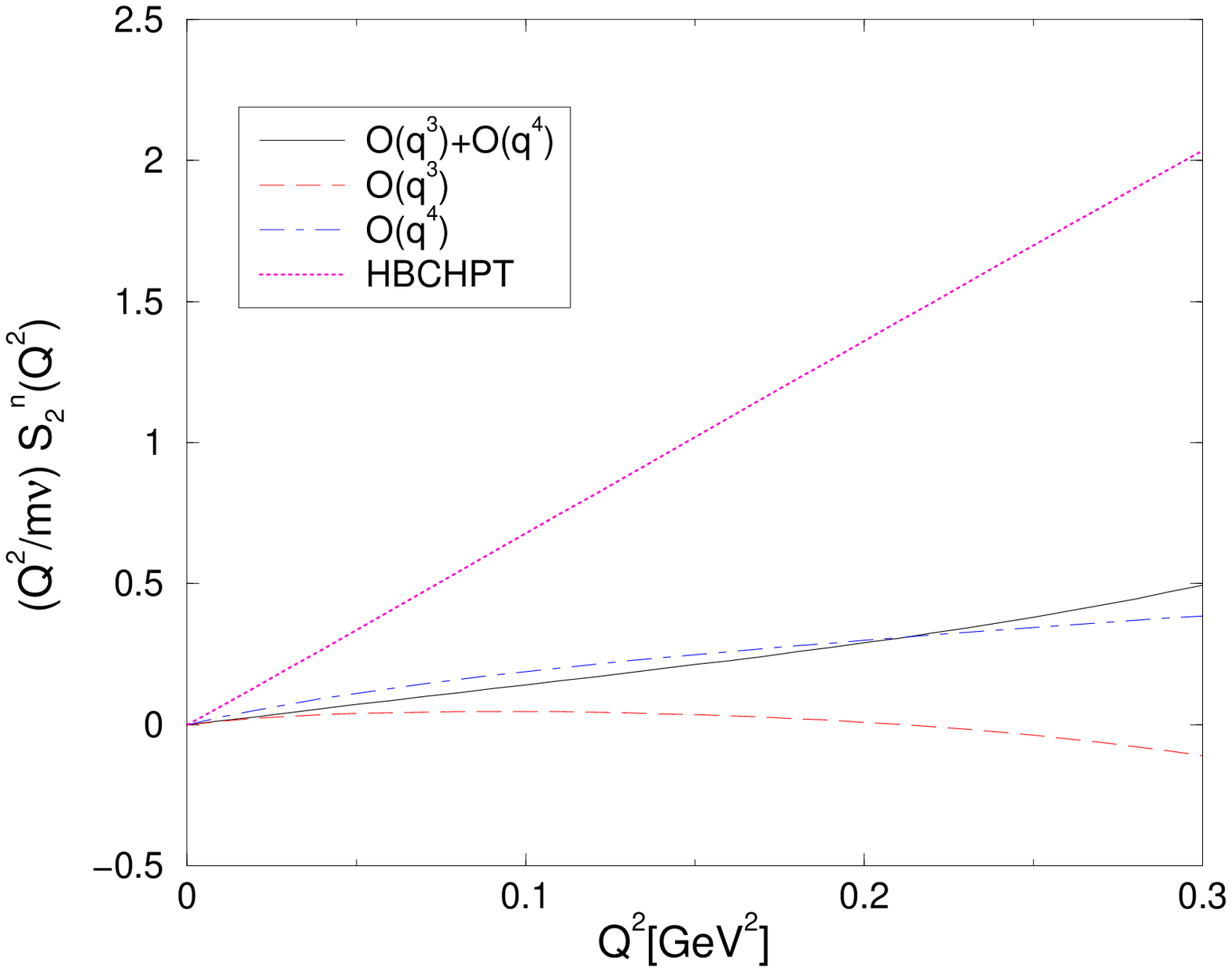,width=.47\textwidth,silent=,clip=}}
\vspace{0.3cm}
\begin{center}
\caption{\label{fig:s2pn}
          Chiral loop contribution to the structure function 
          $(Q^2/m\nu ) \bar{S}_2 (0,Q^2)|_{\nu =0}$ with the 
          elastic contribution subtracted in units of GeV$^{-2}$. 
          The solid (dashed/dot-dashed) line gives the result of
          the present calculation to order $q^4$ (the contribution
          from the third/forth order) in comparison
          to the heavy baryon result of \protect\cite{Ji2} (dotted line).
          Left (right) panel: Proton (neutron).
}
\end{center}
\end{figure}

\begin{figure}[H]
\vspace{.5cm}
\parbox{.49\textwidth}{\epsfig{file= 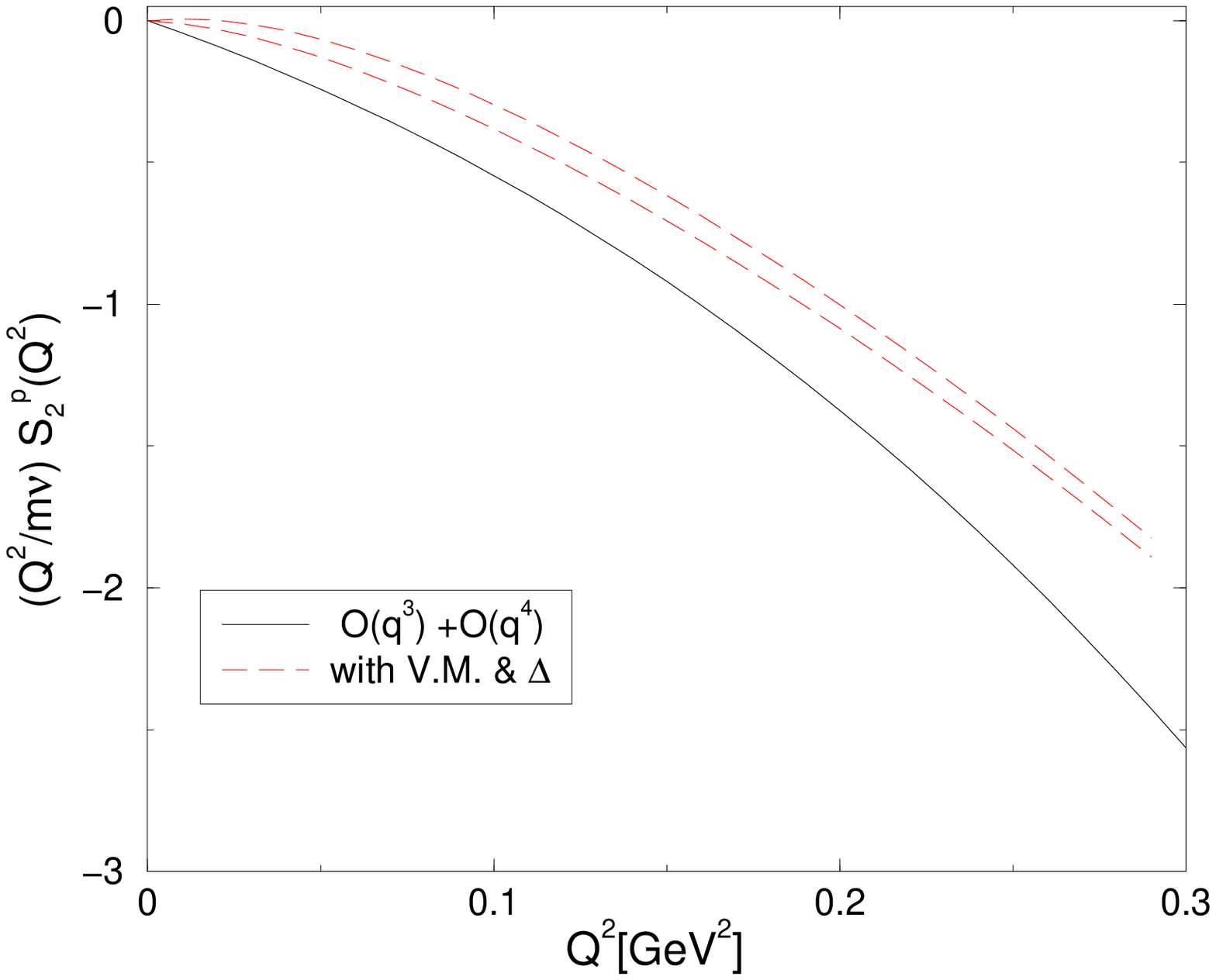,width=.47\textwidth,silent=,clip=}}
\hfill
\parbox{.49\textwidth}{\epsfig{file= 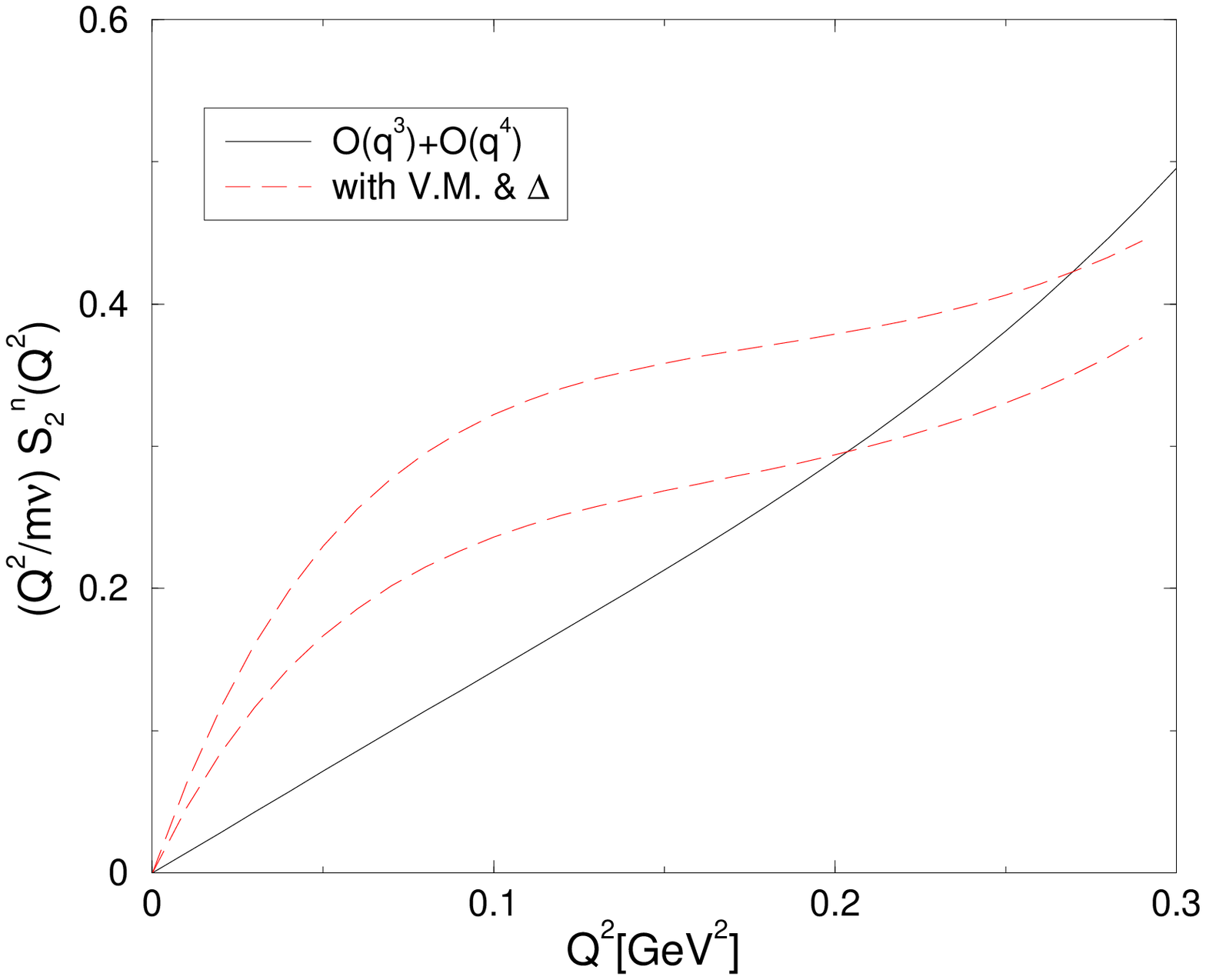,width=.47\textwidth,silent=,clip=}}
\vspace{0.3cm}
\begin{center}
\caption{\label{fig:s2pnr}
          Chiral loop (solid line) and one--loop plus resonance contributions 
          (dashed lines) to the structure function $(Q^2/m\nu ) \bar{S}_2 (0,Q^2)|_{\nu =0}$
          with the elastic contribution subtracted in units of GeV$^{-2}$. 
          Left (right) panel: Proton (neutron).
}
\end{center}
\end{figure}

\begin{figure}[H]
\vspace{0.5cm}
\parbox{.49\textwidth}{\epsfig{file= 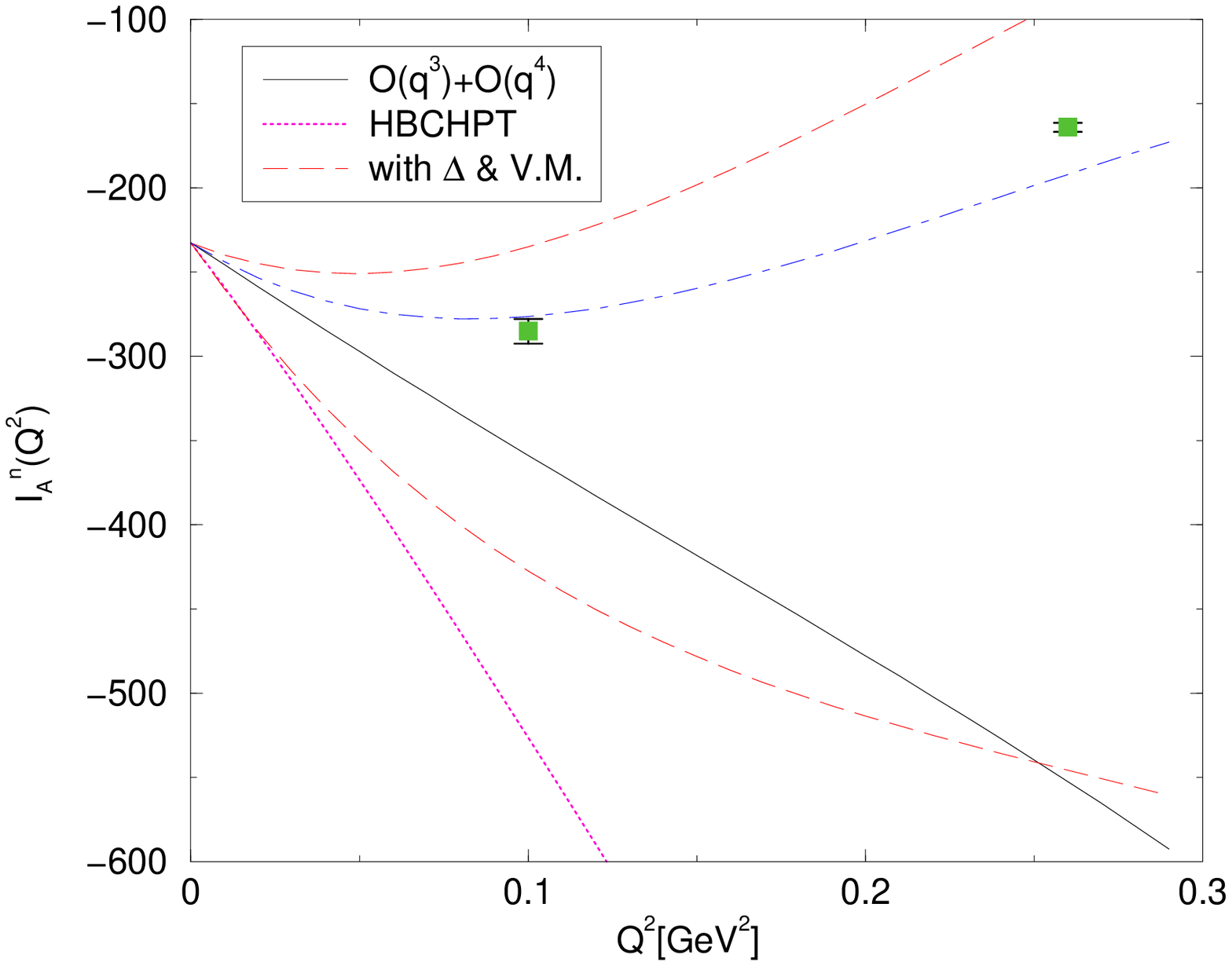,width=.47\textwidth,silent=,clip=}}
\hfill
\parbox{.49\textwidth}{\epsfig{file= 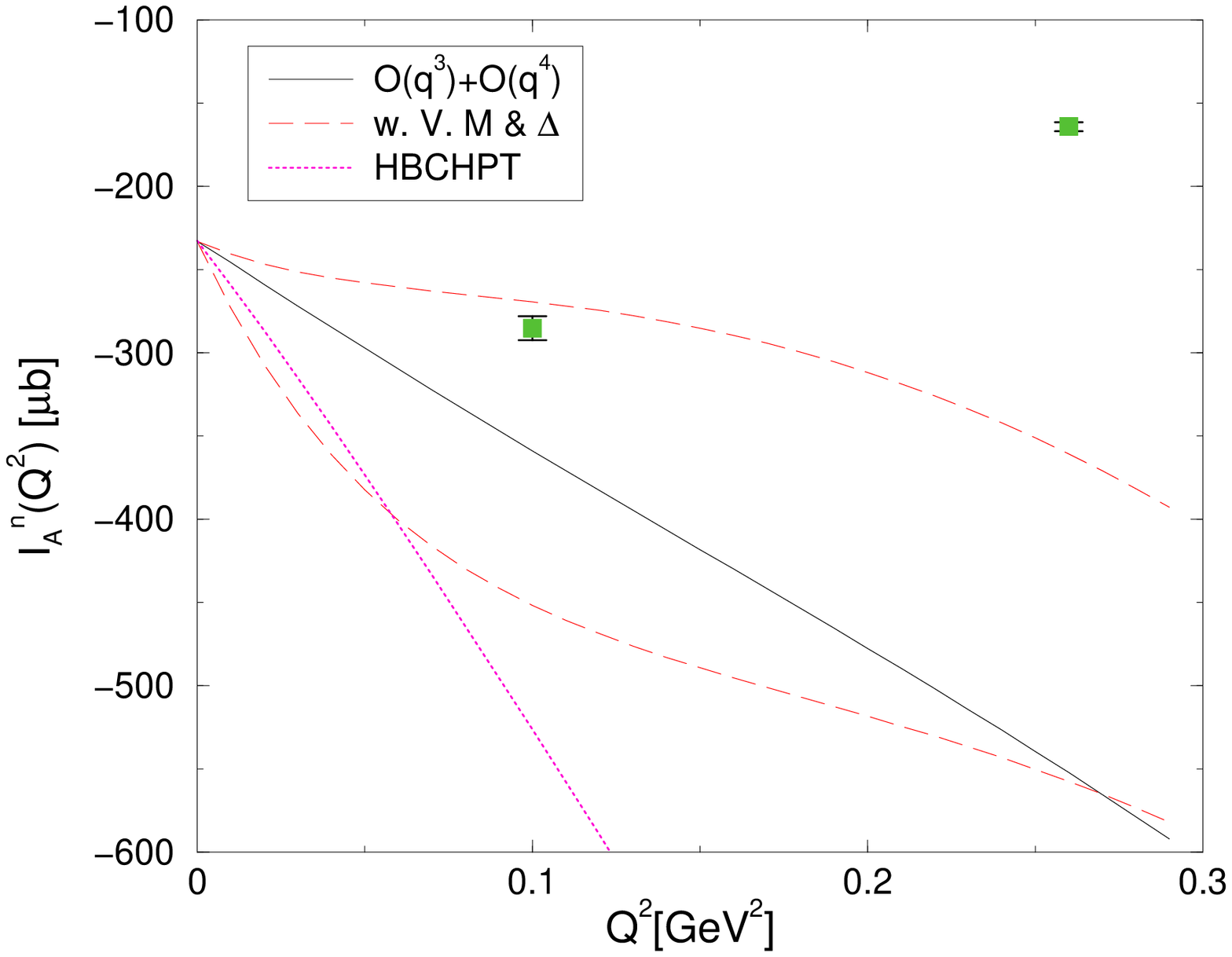,width=.47\textwidth,silent=,clip=}}
\vspace{0.3cm}
\begin{center}
\caption{The integral $I_A (Q^2)$ for the neutron in units of $\mu$b. 
   The solid line gives the
   fourth order result, the dashed lines represent the theoretical uncertainty
   due to variation in the delta parameters as explained in Section~\ref{sec:reso}. 
   For comparison, the HBCHPT result is depicted by the dotted line. 
   The data are from Ref.~\protect\cite{E94}. Note that here we have
   adjusted our normalization to the one used by the experimentalists. Left (right)
   panel: Without (with) $\Delta N\gamma$ form factor. The dot--dashed line in the
   left panel corresponds to the delta parameters $g_2 = 9.5$, $X=Y=-0.725$ as
   discussed in the text. 
\label{fig:IAndat}}
\end{center}
\end{figure}

\begin{figure}[H]
\vspace{1.5cm}
\centerline{
\epsfysize=7in
\epsffile{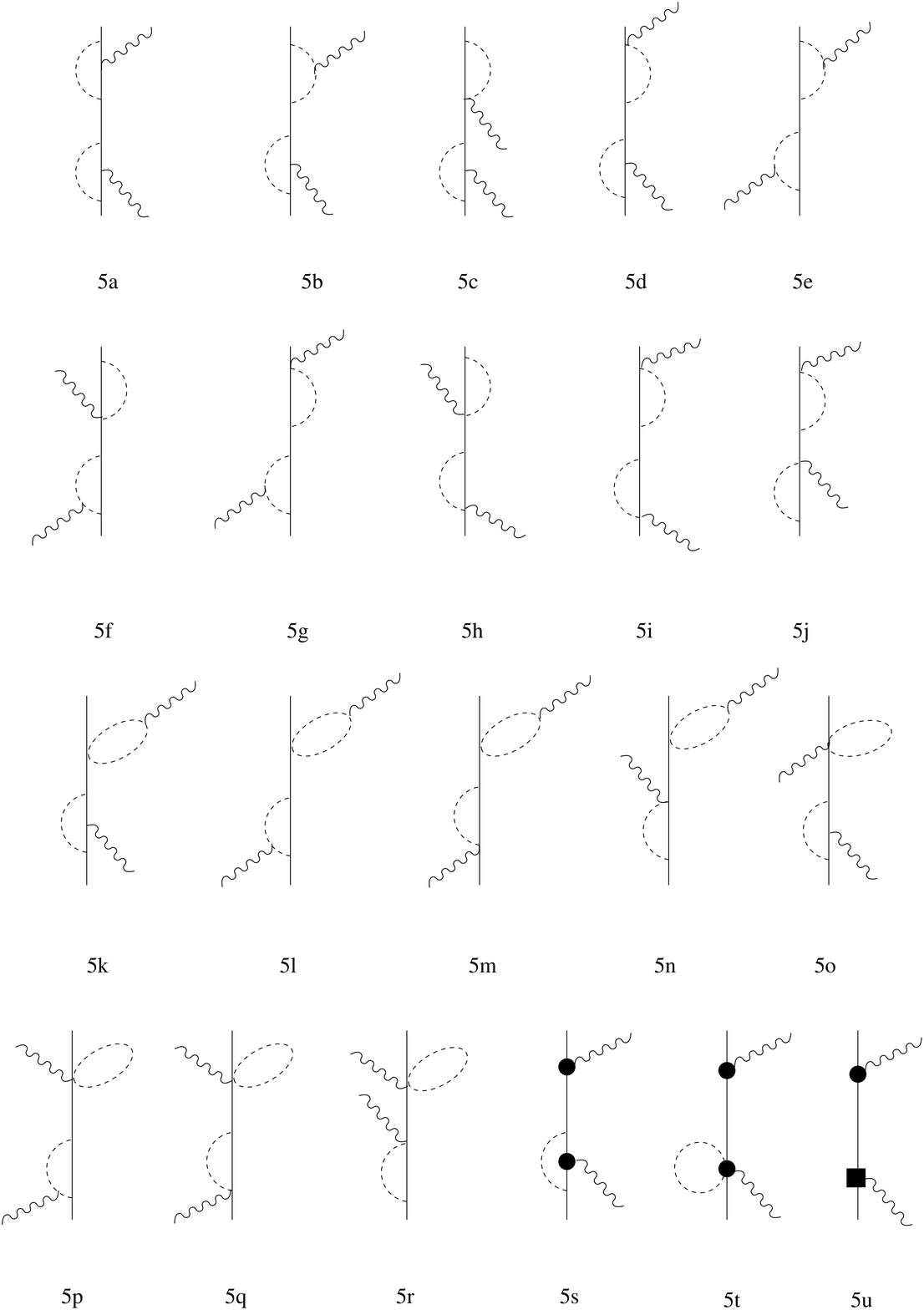}
}
\vspace{1cm}
\begin{center}
\caption{Topologically different pole diagrams for V$^2$CS off
the neutron contributing at fifth order to the combination of
structure functions $S_1 (\nu,Q^2) - (Q^2/m \nu ) \, S_2 (\nu,Q^2)$. 
Solid, dashed and wiggly lines denote nucleons, pions and photons, in order. 
The black filled circle (square) denotes an insertion from the dimension 
two (four) effective 
Lagrangian. All other insertions are  from the leading order (dimension one) 
effective Lagrangian. Crossed diagrams are not shown.
\label{fig:diag5}}
\end{center}
\end{figure}

\begin{figure}[H]
\vspace{0.5cm}
\centerline{
\epsfysize=2.72in
\epsffile{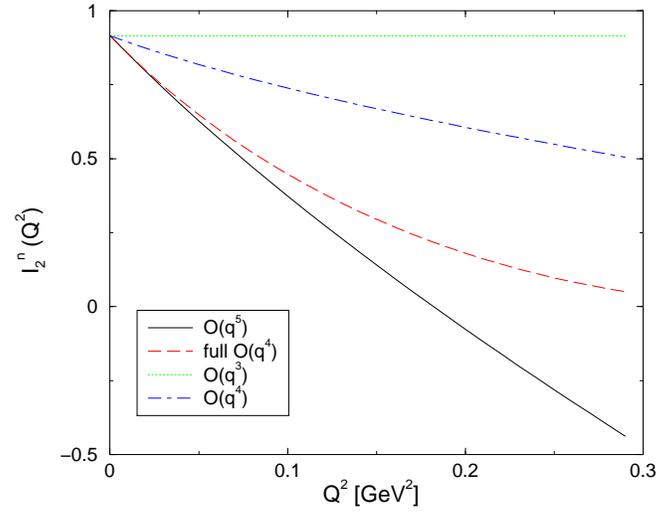}
}
\vspace{0.3cm}
\begin{center}
\caption{The integral $I_2 (Q^2)$ for the neutron.
   The dashed lines represents the full fourth order results (that is
   the form factors taken to fourth order). The
   solid/dot-dashed/dotted line gives the fifth/fourth/third order
   result (i.e. the product of the form factors taken to that order).
\label{fig:I2n5}}
\end{center}
\end{figure}

\medskip\noindent
The pertinent results of this investigation can be summarized as
follows:

\begin{itemize}
\item[1)]We have studied spin--dependent doubly virtual Compton scattering 
off nucleons to fourth order (one loop) in a Lorentz--invariant formulation 
of baryon chiral perturbation theory. At this order in the chiral expansion, 
no unknown low--energy constants appear and thus we obtain parameter--free
predictions for the spin structure functions $\bar{S}_{1,2} (\nu , Q^2)$, 
where the bar signals the subtraction of the elastic intermediate state. 
These two V$^2$CS structure functions serve as the fundamental building blocks 
from which all other results can be derived. 
\item[2)]We have studied the corresponding sum rules (the generalization
of the Gerasimov--Drell--Hearn sum rule to finite photon virtuality and 
the Burkhardt--Cottingham sum rule) and their
moments, which can be obtained by expanding the structure functions
at low energies $\nu$, cf. Eqs.~(\ref{Snu},\ref{S1SR},\ref{S2SR}).
\item[3)]We have estimated the contribution of $\Delta (1232)$ Born graphs and
from vector meson contributions as described in Section~\ref{sec:reso}.
The delta plays an important role in the nucleon spin sector and thus
an extension of our work including $\pi \Delta$ loops in a systematic manner
is called for.
\item[4)]Results for  $\bar S_1^{p,n} (0, Q^2)$ and the first moment 
$\Gamma_1^{p,n} (Q^2)$ 
as well as for $\Gamma_1^{p-n} (Q^2)$ were already given and discussed
in \cite{BHMdhg}. We note that the prediction for $\Gamma_1^{p-n} (Q^2)$
is free of spin--3/2 effects and in rough agreement with preliminary
data from Jefferson Laboratory \cite{RdV}. Predictions for the 
structure functions $\bar S_1^{p,n} (0,Q^2)$ and for the moment $\Gamma_1^p (Q^2)$ 
including resonance contributions,  as well as for the chiral expansion of the second
moments $\bar S_1^{(2)p,n} (0,Q^2)$ are shown in Fig.~\ref{fig:s1pn},
Fig.~\ref{fig:gamma1p} and
Fig.~\ref{fig:s12pn}, respectively. 
\item[5)]The one--loop results for the second structure function are
shown in Fig.~\ref{fig:s2pn}  for the proton and the neutron. They come
out sizeably different from the heavy baryon result for the same reasons
as already discussed in \cite{BHMdhg} in connection with the first structure
function. The delta Born and vector meson contributions have also been
evaluated, see Fig.~\ref{fig:s2pnr}.
\item[6)]The chiral prediction for a linear combination of first moments of $g_1$ and 
$g_2$ commonly denoted as $I_A^n (Q^2)$ supplemented with 
resonance contributions (mostly from the delta) agrees with recent
data from Jefferson Laboratory \cite{E94} for virtualities $Q^2 \lesssim 0.3\,$GeV$^2$,
see Fig.~\ref{fig:IAndat}.  However, if one includes a
$\Delta N \gamma$ vertex function as suggested by older electroproduction data
at higher energies, one is not able to describe
the curvature of the data as the virtuality increases,
see also Fig.~\ref{fig:IAndat}.
\item[7)] The  Burkhardt--Cottingham sum rule is fulfilled
within the Lorentz--invariant approach used here. However, truncating the 
product of form factors at fourth order does not give a very accurate
representation of the full result, which can be obtained using the form factors
from dispersion theory or infrared regularized baryon CHPT 
supplemented by vector mesons \cite{KM}.
\item[8)]We have calculated  the forward spin polarizability
  $\gamma_0 (Q^2)$ (which corresponds to a 
linear combination of the second moments of
$g_1$ and $g_2$) at zero and finite
photon virtuality to ${\cal O}(p^4)$ in relativistic baryon CHPT for 
the proton and the neutron.  To obtain a reliable
prediction for the chiral corrections to these spin polarizabilities at $Q^2=0$ in the
heavy baryon approach, we find that 
one has to account (at least) for terms of ${\cal O}(M_\pi)$, 
i.e. go to N$^4$LO in the 
calculation.  However, we note that even our 
relativistic one-loop result, 
when supplanted with resonance model delta Born term and 
vector meson contributions, still does not agree with the data \cite{Ahrens}.
We have also isolated the reason for the bad convergence, it rests entirely on
the slow convergence of the pole (Born) graphs. 
\item[9)] We have calculated  the longitudinal--transverse spin polarizability 
$\delta_0(Q^2)$ (which corresponds to a linear combination of the second moment of 
$g_1$ and the first moment of $g_2$) at zero and finite photon virtuality for proton
and neutron. When performing a chiral expansion on the full relativistic results, 
including terms up--to--and--including  ${\cal O}(M_\pi)$, one finds a good 
convergence for these quantities at the photon point in the $1/m$ expansion. This 
observation and the absence of 
large delta (Born-term) contributions leads to the hope that $\delta_0(Q^2)$ 
can serve as
 a testing ground of the chiral dynamics of QCD which might be better suited than the 
forward spin polarizabilities of the nucleon.
\item[10)] We have calculated the fifth order contributions from the pole (Born)
graphs for a particular combination of the structure functions of 
the neutron shown in Fig.~\ref{fig:diag5}. 
The complete fifth order result for the neutron BC sum rule is in
good agreement with the product of the form factors for 
$Q^2 \lesssim 0.15\,$GeV$^2$, cf. Fig.~\ref{fig:I2n5}. The partial
fifth order contribution to the neutron's spin polarizability is
non--negligible and negative, indicating that these terms might be
the origin for explaining the discrepancy to the data in case of the proton. 
\end{itemize}

\medskip\noindent
The material in this paper is organized as follows. Section~\ref{sec:formalism}
contains the basic formalism for spin-dependent doubly virtual Compton scattering
off nucleons, its relation to pion electroproduction and the derivation of the
pertinent sum rules and their moments. In section~\ref{sec:chpt}, the theory 
underlying our calculations is exposed together with the chiral expansion of the
spin--dependent structure functions. We also show how we estimate various
resonance contributions. The results are presented and discussed in 
section~\ref{sec:res}. Some technicalities are relegated to the appendix.


\section{Basic formalism}
\def\theequation{\arabic{section}.\arabic{equation}}
\setcounter{equation}{0}
\label{sec:formalism}

This section contains the basic formalism on spin-dependent doubly
virtual Compton scattering off nucleons, its relation to 
inelastic electroproduction
and the derivation of the corresponding sum rules. The reader familiar
with these concepts might directly proceed to section~\ref{sec:chpt}.

\subsection{Doubly virtual Compton scattering}
\label{sec:V2CS}
We consider spin--dependent doubly virtual Compton scattering (V$^2$CS) 
off nucleons (neutrons or protons) in forward direction,
that is the reaction 
\beq
\gamma^\star (q,\epsilon) + N(p,s) \to \gamma^\star (q,\epsilon ')+ N(p,s^\prime)~,
\eeq
with $q~(p)$ the virtual photon (nucleon) four--momentum, $s~(s^\prime)$
the nucleon spin (polarization) and $\epsilon\, (\epsilon ')$ the
polarization four--vector of the incoming (outgoing) photon.
It is common to express the spin amplitude of V$^2$CS in terms of two structure 
functions, called $S_1(\nu,Q^2)$ and $S_2(\nu,Q^2)$, via
\begin{eqnarray}
T^{[\mu\nu]}(p,q,s)=-i\;\frac{m}{N}\,\epsilon^{\mu\nu\alpha\beta}q_\alpha\left\{s_\beta
\, S_1(\nu,Q^2)+\left[p\cdot q\, s_\beta-s\cdot q\;p_\beta\right]
\frac{S_2(\nu,Q^2)}{m^2}
\right\},
\end{eqnarray}
where $s^\mu$ denotes a spin-polarization four-vector, $m$ is the nucleon mass, 
$\epsilon^{\mu\nu\alpha\beta}$ the totally antisymmetric Levi--Civita tensor with 
$\epsilon^{0123}=1$,
$\nu=p \cdot q/ m $ the energy transfer and $Q^2 = -q^2 \ge 0$ the (negative of
the) photon virtuality. $N$ is a convention dependent normalization factor. Following 
Ref.\cite{BHMdhg} we set $S_1(0,0)=-e^2 \,\kappa^2/ m^2$. With the spinor normalization
$\bar{u}u=1$ typically used in baryon CHPT, we find $N=2m$, 
which we will utilize throughout this  work.
Note that while $S_1 (\nu,Q^2)$ is even under crossing 
$\nu \leftrightarrow -\nu$,  the structure function $S_2(\nu,Q^2)$ is odd.

\medskip\noindent
In the rest frame of the nucleon the forward virtual Compton tensor in 
the Coulomb gauge $\epsilon_0=\epsilon^\prime_0=0$  simplifies to
\beqa
T^{[ij]}_{\rm rest} (\nu,Q^2)=\frac{i}{2}\;
\epsilon_{ijk}\chi^\dagger\left\{
\nu\,\sigma^kS_1(\nu,Q^2)+\left[\nu^2\sigma^k-\sigma\cdot q\,q^k\right]
\frac{S_2(\nu,Q^2)}{m}\right\}\chi~,
\eeqa
where $\nu$ now corresponds to the photon energy in the laboratory system,
the $\sigma_k$ $(k=1,2,3)$ are the Pauli spin--matrices 
and $\chi$ is a two-component spinor.  Utilizing the identity
\begin{eqnarray}
\vec{\sigma}\cdot\vec{q}\,\vec{\epsilon}^{\,\prime}\cdot(\vec{\epsilon}
\times\vec{q})=\left(\nu^2+Q^2\right)\left[\vec{\sigma}
\cdot(\vec{\epsilon}^{\,\prime}\times\vec{\epsilon})-\left(\vec{\sigma}
\cdot(\vec{\epsilon}^{\,\prime}\times\hat{q})\,
\vec{\epsilon}\cdot\hat{q}-\vec{\sigma}\cdot(\vec{\epsilon}\times\hat{q})\,
\vec{\epsilon}^{\,\prime}\cdot\hat{q}\right)\right]
\end{eqnarray} 
we finally arrive at the V$^2$CS forward matrix element
\begin{eqnarray}
\epsilon^\prime\cdot T\cdot\epsilon|_{\rm rest}
&=&\frac{1}{2m}\, \chi^\dagger\left\{i\vec{\sigma}\cdot
\left(\vec{\epsilon}^{\,\prime}\times\vec{\epsilon}\right)\left[m\nu 
S_1(\nu,Q^2)-Q^2S_2(\nu,Q^2)\right]\right.\nonumber\\
& &\phantom{\frac{1}{N}\;\chi^\dagger}\left.
+i\left[\vec{\sigma}\cdot(\vec{\epsilon}^{\,\prime}\times\hat{q})\,
\vec{\epsilon}\cdot\hat{q}-\vec{\sigma}\cdot(\vec{\epsilon}\times\hat{q})\,
\vec{\epsilon}^{\,\prime}\cdot\hat{q}\right]\left(\nu^2+Q^2\right)S_2(\nu,Q^2)
\right\}\chi\; ,\label{match}
\end{eqnarray} 
in the form which is best suited for our calculations in the framework
of relativistic baryon chiral perturbation theory.
The Compton amplitudes $S_{1,2} (\nu, Q^2)$ are amenable to a chiral 
expansion as will be detailed in section~\ref{sec:chpt}. We remark that
in what follows, we will  mostly be concerned with the reduced amplitudes 
\beq\label{red}
\bar{S}_i (\nu,Q^2) =
S_i (\nu,Q^2) - S_i^{\rm el}  (\nu,Q^2)~,
\eeq 
i.e. the Compton amplitudes with the contribution from the elastic intermediate 
state subtracted. More precisely, these are the contributions from the single nucleon
exchange (pole) terms with the corresponding vertices given in terms of the electromagnetic
form factors. Only the non-pole parts of the corresponding diagrams contribute to the
nucleon spin structure as discussed in more detail below. We note that
for the calculations to follow we mostly work at $\nu = 0$ (and 
therefore often do not make explicit the energy dependence of certain
quantities). In fact, for the calculations we make use of  crossing
symmetry, which allows to write the relativistic V$^2$CS forward matrix element
in terms of two invariant functions $A(\nu,Q^2)$ and $B(\nu,Q^2)$ as\footnote{A similar
method was used e.g. in Refs.~\cite{BKMpol,BKMdhg,EKPW}.}:
\beqa\label{AB}
\epsilon^\prime\cdot T\cdot\epsilon|_{\rm rest}
&=&\chi^\dagger\left\{i\vec{\sigma}\cdot
\left(\vec{\epsilon}^{\,\prime}\times\vec{\epsilon}\right)\left[A(s,Q^2)
-A(2m^2-2Q^2-s,Q^2)\right]\right.\nonumber\\
&+&\left.
 i\left[\vec{\sigma}\cdot(\vec{\epsilon}^{\,\prime}\times\hat{q})\,
 \vec{\epsilon}\cdot\hat{q}-\vec{\sigma}\cdot(\vec{\epsilon}\times\hat{q})\,
 \vec{\epsilon}^{\,\prime}\cdot\hat{q}\right] \, 
 |\vec{q}\,|^2 
 \, \left[B(s,Q^2)
 -B(2m^2-2Q^2-s,Q^2)\right]\right\}\chi\; ,\label{TAB} \nonumber \\ &&
 \eeqa
with $s=(p+q)^2$ the standard Mandelstam variable, which is related to $\nu$
via $\nu = (s- m^2 + Q^2) / 2m$.
Written in term of $\nu$, crossing symmetry means that in this equation 
the difference of a quantity taken at $\nu$
with the same quantity taken at $-\nu$ appears.
Since we consider quantities at  $\nu=0$ i.e. at $s=m^2+q^2$, all of
these can be obtained as derivatives of order $n \ge 1$ of the functions $A$ and $B$. 
More precisely, we give two examples to illustrate this method,
\beqa
\bar S_1(0,Q^2)-{Q^2 \over m} \lim_{\nu \to 0}  {\bar {\underline S}_2(0,Q^2)
\over \nu} 
&=& 4\, {d A(\nu, Q^2) \over d\nu} \bigg|_{\nu=0}~, \\
Q^2 \lim_{\nu \to 0}  {\bar {\underline S}_2(0,Q^2) \over \nu}
&=& 4\, {d B(\nu, Q^2) \over d\nu} \bigg|_{\nu=0}~, 
\eeqa
where $\nu \bar {\underline S}_2(\nu,Q^2)=\nu \bar S_2(\nu,Q^2) -
\nu\bar S_2(\nu,Q^2)\big|_{\nu=0}\,$.

\subsection{Relation to inelastic electroproduction off nucleons}
Because of unitarity, there is a basic connection between spin  
structure functions in  V$^2$CS and the ones probed in inelastic  
electroproduction experiments. This is simply related to the fact that
the imaginary part of the Compton tensor is given in terms of nucleon
plus meson states, the lowest one being the pion--nucleon state.
The differential cross section for exclusive electroproduction can
be expressed in terms of four virtual photoabsorption cross sections
$\sigma_{T}(\nu,Q^2)$, $\sigma_{L}(\nu,Q^2)$, $\sigma_{LT}^\prime(\nu,Q^2)$ 
and $\sigma_{TT}^\prime(\nu,Q^2)$. The transverse, $\sigma_T$, and 
the transverse--transverse, $\sigma_{TT}^\prime$, cross sections
can be related to  the total photoabsorption cross sections
$\sigma_{1/2}$
and $\sigma_{3/2}$ corresponding to the excitation of intermediate states
with spin projection 3/2 and 1/2, respectively:
\beqa\label{sig13}
\sigma_T={1 \over 2}\left(\sigma_{3/2} + \sigma_{1/2}\right)~, \cr \cr
\sigma_{TT}^\prime={1 \over 2}\left(\sigma_{3/2} - \sigma_{1/2}\right)~.
\eeqa
The four virtual photoabsorption cross sections are related to the 
quark structure functions  $F_1$, $F_2$, $G_1$ and $G_2$ (that
depend on $\nu$ and $Q^2$) measured in
(un)polarized deep inelastic electron scattering off nucleons via:
\beqa\label{S}
\sigma_T(\nu,Q^2)&=&{4\pi^2\alpha \over m K} F_1(\nu,Q^2) ~,
\\
\sigma_L(\nu,Q^2)&=&{4\pi^2\alpha \over  K}\Bigl[{F_2(\nu,Q^2) \over \nu}(1+\gamma^2)-
{F_1(\nu,Q^2) \over m}\Bigr] ~,
\\\label{S1}
\sigma_{LT}^\prime(\nu,Q^2)&=&-{4\pi^2\alpha \over  K} \, \gamma\; \nu\; (G_1(\nu,Q^2) 
+{\nu \over m} G_2(\nu,Q^2)) ~,\\
\sigma_{TT}^\prime(\nu,Q^2)&=&-{4\pi^2\alpha \over K} \, \nu\; (G_1(\nu,Q^2) -
{\nu \over m} \gamma^2 G_2(\nu,Q^2)) ~,\label{STTp}
\eeqa
where $\nu =E-E'$ is the virtual photon energy 
in the lab frame, $\alpha = e^2/4\pi = 1/137.036$ the fine structure constant,
$\gamma=Q/ \nu$ and $K$ a flux factor. For more details on the electroproduction
formalism, see e.g. Ref.\cite{Drechsel}.

\subsection{Sum rules and moments}
\label{sec:sumrules}
Let us first consider the real photon case with $Q^2=0$. At low energies one expands 
the coefficient 
\beq
2\,m\,g(\nu)= \lim_{Q^2\to 0} m\nu S_1(\nu,Q^2)-Q^2S_2(\nu,Q^2)
\eeq 
in front of the spin-dependent operator
$\vec{\sigma}\cdot \left(\vec{\epsilon}^{\,\prime}\times\vec{\epsilon}\,\right)$ in
powers of the energy transfer,
\beqa\label{E}
g(\nu) = -{e^2 \kappa^2 \over 2 m^2} \nu +4 \pi \; \gamma_0\; \nu^3
+ {\cal O}(\nu^5)~,
\eeqa
where the first term proportional to the nucleon anomalous magnetic
moment squared ($\sim \kappa^2$) is given by the venerable low--energy theorem  (LET) of 
Low, Gell-Mann and Goldberger \cite{LGG} 
and $\gamma_0$ is the forward spin polarizability.
The imaginary part of $g(\nu)$ can be related via the optical theorem to  
$\sigma_{1/2}$ and $\sigma_{3/2}$,
\beqa\label{O}
{\rm Im} \; g(\nu)={\nu \over 2} \left(\sigma_{1/2}(\nu,Q^2=0)
 -\sigma_{3/2}(\nu,Q^2=0)\right)
\eeqa
Writing an unsubtracted dispersion relation for $g(\nu)$ one obtains
using Eq.(\ref{E}) and (\ref{O}) the celebrated Drell--Hearn--Gerasimov 
sum rule \cite{GDH} (we give here the generic form without specifying whether
we mean the neutron or the proton):
\beqa 
-{2\;\pi^2\;\alpha\;\kappa^2 \over m^2} =\int {d\nu \over \nu}\, \left[
\sigma_{3/2} - \sigma_{1/2} \right]~.
\eeqa
The value of the LHS of the sum rule is $-204 \,\mu$b for the proton and 
$-232 \,\mu$b  for the neutron. This sum rule is presently under active experimental
investigation at MAMI, ELSA, GRAAL, CEBAF and Spring-8. For first results, 
see~\cite{GDHexp}.

\medskip\noindent
The DHG sum rule has been  generalized to finite  $Q^2$ \cite{AJL}. 
It is most natural to connect such a generalization of the sum rule to 
the Bjorken sum rule in deep inelastic scattering. One thus defines 
\beqa
I_1(Q^2)={ 2 m^2 \over Q^2} \int_0^{x_0} g_1(x,Q^2) dx~, 
\eeqa
with $x_0$ corresponding to the pion production threshold.
Here, $G_1= g_1 / (m \nu)$ can be related to $\sigma_{3/2}$ and 
$\sigma_{1/2}$ via Eqs.(\ref{sig13},\ref{STTp}) 
and $x=Q^2/2m\nu$ is the standard scaling variable.
At zero virtuality, one recovers the DHG sum rule, i.e. $I_1(0)= -\kappa^2/4$.
Similarly, there exists a sum rule for the second spin structure function, $g_2$,
the so--called  Burkhardt--Cottingham (BC) sum rule \cite{BC}:
\beqa\label{BCsr}
I_2(Q^2)&=&{2 m^2  \over Q^2} \int_0^{x_0}  g_2(x,Q^2) dx
\\
        &=&{1 \over 4} {G_M(Q^2)-G_E(Q^2) \over (1+{Q^2 \over 4m^2})}
        G_M(Q^2) = {1 \over 4} F_2 (Q^2) \, \left( F_1 (Q^2) + F_2
          (Q^2) \right)~,
\eeqa
where $G_E (Q^2)$ and $G_M (Q^2)$  are the electric and the magnetic nucleon 
Sachs form factor, respectively, $F_1 (Q^2)$ and $F_2 (Q^2)$ are the
Dirac and the Pauli form factor, in order, and $G_2= g_2 / \nu^2$. 
We note that in a Lorentz-invariant calculation as performed here, the
Dirac and Pauli form factors appear naturally, while in the HBCHPT
approach the Sachs form factors are the more appropriate quantities \cite{BKKM}.
This sum rule assumes that 
$\sigma_{LT}^\prime$ vanishes faster than $1/\nu$ at large $\nu$. 
Other sum rules have been defined in the literature, see e.g. \cite{BKMdhg,Drechsel}:
\beqa
I_A(Q^2) &=& { 2 m^2 \over Q^2} \int_0^{x_0} (g_1(x,Q^2) - \gamma^2 g_2(x,Q^2)) dx~,
\\
 I_C(Q^2) &=& { 2 m^2 \over Q^2} \int_0^{x_0}
{1 \over (1-x)} (g_1(x,Q^2) - \gamma^2 g_2(x,Q^2) ) dx   ~.
\eeqa

\medskip\noindent
All these sum rules can be written in terms of the structure functions 
${S}_1(\nu,Q^2),\; {S}_2(\nu,Q^2)$ of V$^2$CS using the following dispersion 
relations :
\begin{eqnarray}
{S}_1(\nu,Q^2)&=&4e^2\int_{Q^2/2m}^\infty\frac{d\nu^\prime\nu^\prime 
               G_1(\nu^\prime,Q^2)}{\nu^{\prime 2}-\nu^2}~,\nonumber\\
\nu {S}_2(\nu,Q^2)&=&4e^2 \int_{Q^2/2m}^\infty\frac{d\nu^\prime \nu^{2} 
               G_2(\nu^\prime,Q^2)}{\nu^{\prime 2}-\nu^2}\; . 
               \label{iofferules}
\end{eqnarray}
where use has been made of the optical theorem,
\beq
{\rm Im}~{S}_i (\nu , Q^2)  = 2\pi~ G_i (\nu , Q^2) ~, \quad (i = 1,2)~.
\eeq
Expanding the structure functions at low energies $\nu$, that is around 
$\nu = 0$ (note that we now consider the structure functions with the elastic
intermediate state subtracted),
\begin{eqnarray}\label{Snu}
\bar{S}_1(\nu,Q^2)&=&\bar{S}_1^{(0)}(0,Q^2)+\nu^2\,\bar{S}_1^{(2)}(0,Q^2)+
                     \nu^4\,\bar{S}_1^{(4)}(0,Q^2)+\dots~, \nonumber\\
\nu \bar{S}_2(\nu,Q^2)&=&\bar{S}_2^{(-1)}(0,Q^2)+
\nu^2\,\bar{S}_2^{(1)}(0,Q^2)+\nu^4\,\bar{S}_2^{(3)}(0,Q^2)+
                     \nu^5\,\bar{S}_2^{(5)}(0,Q^2)+\dots \; ,
\end{eqnarray}
where the ellipsis denote terms of higher order in $\nu$, 
one  obtains the following  set of sum rules for $\bar{S}_1(0,Q^2)$:
\begin{eqnarray}\label{S1SR}
\bar{S}_1^{(0)}(0,Q^2)&=&4e^2\int_{\nu_0}^\infty\frac{d\nu^\prime 
               G_1(\nu^\prime,Q^2)}{\nu^{\prime}}={4e^2 \over m^2} I_1(Q^2)~,\\
\bar{S}_1^{(2i)}(0,Q^2)&=&4e^2\int_{\nu_0}^\infty\frac{d\nu^\prime 
               G_1(\nu^\prime,Q^2)}{\nu^{\prime (2i+1)}} \quad (i=1,2,3,\dots)~,
\end{eqnarray}
noting that  $\bar{S}_1^{(0)}(0,Q^2)$ is directly related to  $I_1(Q^2) $.
Note that the often used first moment $\Gamma_1 (Q^2)$ is related to
$I_1 (Q^2)$ via
\beq
\Gamma_1 (Q^2) = \frac{Q^2}{2m^2} \, {I}_1 (Q^2)~.
\eeq
Likewise, $\bar{S}_2^{(i)} (0,Q^2)$ satisfies:
\begin{eqnarray}\label{S2SR}
\bar{S}_2^{(-1)}(0,Q^2)&=&4e^2\int_{\nu_0}^\infty d\nu^\prime 
               G_2(\nu^\prime,Q^2)={4 \over m^2} I_2(Q^2)~,\\
\bar{S}_2^{(i)}(0,Q^2)&=&4e^2\int_{\nu_0}^\infty\frac{d\nu^\prime 
               G_2(\nu^\prime,Q^2)}{\nu^{\prime (i+1)}} \quad (i = 1,3,5,\ldots)~,
\end{eqnarray}
where $\bar{S}_2^{(-1)}(0,Q^2)$ is related to  $I_2$.
In order to calculate $I_C$ it is necessary to  
define a new quantity $\bar{S}_R$ similar to  $\bar{S}_1-Q^2\bar{S}_2/\nu$,
but having only right hand cuts:
\beqa
\bar{S}_R(\nu,Q^2)=4e^2 \int_{\nu_0}^\infty d\nu^\prime {G_1(\nu^\prime,Q^2)
 -Q^2 G_2(\nu^\prime,Q^2)/\nu \over \nu^\prime-\nu}~,
\eeqa
so that the following relations hold:
\beqa
I_A(Q^2)&=&{m^2 \over 4} \left(\bar{S}_1^{(0)}(0,Q^2)-Q^2\bar{S}_2^{(1)}(0,Q^2)\right)~,
\\
I_C(Q^2)&=&{m^2 \over 4} \bar{S}_R(Q^2/2m,Q^2)~.
\eeqa

\medskip\noindent
In Eq.(\ref{E})  the forward spin polarizability appears \cite{BKKM}, which is defined
in terms of the transverse--transverse photoabsorption cross section as
\beqa
\gamma_0 ={1 \over 4 \pi^2} \int {d\nu\over\nu^3}
(1-x)(\sigma_{1/2}(\nu,0) - \sigma_{3/2}(\nu,0))~.
\eeqa
Note that $x$ is trivially zero at vanishing virtuality in this equation so that
one recovers the usual dispersion relation for $\gamma_0$.
One can define a similar quantity involving  $\sigma^\prime_{LT}$ instead
of $\sigma_{TT}^\prime$, the so--called longitudinal-transverse polarizability,
denoted $\delta_0$ in what follows,
\beqa
\delta_0 ={1 \over 2 \pi^2} \int {d\nu\over\nu^3}
(1-x) \lim_{Q^2 \to 0}
\biggl({\nu\over Q}(\sigma_{1/2}(\nu,0)\biggr)~.
\eeqa
These two quantities can be generalized to finite $Q^2$. 
The quantity $x$ was in fact  introduced in order to get straightforward
relations between $\gamma_0(Q^2)$, $\delta_0(Q^2)$ and the 
$\bar{S}_i  (\nu , Q^2)$. One has:
\beqa
\gamma_0(Q^2)&=&{1 \over 8\pi}\left(\bar{S}_1^{(2)}(0,Q^2)
- \frac{Q^2}{m}\bar{S}_2^{(3)}(0,Q^2)\right)~,
\\
\delta_0(Q^2)&=&{1 \over 8\pi}\left((\bar{S}_1^{(2)}(0,Q^2)
+\bar{S}_2^{(1)}(0,Q^2)\right)~. 
\eeqa

\section{Chiral expansion of the structure functions}
\def\theequation{\arabic{section}.\arabic{equation}}
\setcounter{equation}{0}
\label{sec:chpt}

Our calculations are based on an  effective chiral pion--nucleon Lagrangian
in the presence of external sources (like e.g. photons)  supplemented
by a power counting in terms of quark (meson) masses and small external momenta.
Its generic form consists of a string of terms with increasing chiral dimension,
\beq\label{L}
{\cal L}_{\rm eff} =
{\cal L}_{\pi N}^{(1)} + {\cal L}_{\pi N}^{(2)} + {\cal L}_{\pi N}^{(3)} +   
{\cal L}_{\pi N}^{(4)} + {\cal L}_{\pi\pi}^{(2)} + {\cal L}_{\pi\pi}^{(4)}
+ \ldots~.
\eeq
The superscript denotes the power in the genuine small parameter $q$ 
(denoting pion masses and/or external momenta). A complete one--loop (fourth
order) calculation must include all tree level graphs with insertions
from all terms given in Eq.~(\ref{L}) and loop graphs with at most
one insertion from ${\cal L}_{\pi N}^{(2)}$. The complete Lagrangian
to this order is given in \cite{FMMS}. We note that for the case under
consideration the only appearing dimension two low--energy constants
(LECs), called $c_6$ and $c_7$
\cite{BKMrev}, can be fixed from the anomalous magnetic moment of the proton and
of the neutron, respectively. As discussed below, there are no contributions
from ${\cal L}_{\pi N}^{(3,4)}$ for the observables considered here.
Besides these chiral corrections to V$^2$CS, we also estimate various resonance
contributions to the observables, however, we do not have an equally systematic
framework at our disposal for that task. So some future work should be undertaken
to sharpen this part of our calculation by e.g. extending the small scale
expansion of \cite{HHK} to Lorentz--invariant formulation.

\subsection{Essentials of Lorentz-invariant baryon CHPT}

\noindent
Baryon chiral perturbation theory 
is complicated by the fact that the nucleon mass does not vanish in
the chiral limit and thus introduces a new mass scale apart from the
one set by the quark masses. Therefore, any power of the quark masses 
can be generated by chiral loops in the nucleon (baryon) case, spoiling the
one--to--one correspondence between the loop expansion and the one
in the small parameter $q$. One method to overcome this is  the heavy mass
expansion (called heavy baryon chiral perturbation theory, for short HBCHPT) 
where the nucleon  mass is
transformed from the propagator into a string of vertices with
increasing powers of $1/m$. Then, a consistent power counting emerges.
However, this method has the disadvantage
that certain types of diagrams are at odds with strictures from analyticity.
The best example is the so--called triangle graph, which enters e.g. the
scalar form factor or the isovector electromagnetic form factors of the
nucleon.  In a fully relativistic treatment,
such constraints from analyticity are automatically fulfilled. It was recently
argued in~\cite{ET} that relativistic one--loop integrals can be separated
into ``soft'' and ``hard'' parts. While for the former the power counting
as in HBCHPT applies, the contributions from the latter can be absorbed in 
certain LECs. In this way, one can combine the advantages 
of both methods. A more formal and rigorous implementation of such a program 
was given in \cite{BL}. The underlying method  is called 
``infrared regularization''. Any dimensionally regularized
one--loop integral $H$ is split into an infrared singular (called $I$) and 
a regular part (called $R$) by a particular choice of Feynman parameterization, 
\beq
H = I + R ~.
\eeq
Consider first the regular part. If one  chirally  expands the contributions
to $R$, one generates
polynomials in momenta and quark masses. Consequently, to any order, $R$ can
be absorbed in the  LECs of the effective Lagrangian.  On the other hand, the
infrared (IR) singular part  has the same analytical properties as the full
integral $H$ in the low--energy region and its chiral expansion leads to the
non--trivial momentum and quark--mass dependences of CHPT, like e.g. the
chiral logs or fractional powers of the quark masses.
For a typical one--loop integral (like e.g. the nucleon self--energy $\Sigma$)
this splitting can be achieved in the following way (omitting prefactors)
\beqa
\Sigma = \int  \frac{d^dk}{(2\pi)^d} {1 \over AB} 
&=& \int_0^1 dz \int  \frac{d^dk}{(2\pi)^d} {1 \over [(1-z)A+zB]^2}
\nonumber \\
&=& \biggl\{ \int_0^\infty - \int_1^\infty \biggr\} dz
 \int  \frac{d^dk}{(2\pi)^d} {1 \over [(1-z)A+zB]^2} = I + R~,
\eeqa
with $A=M_\pi^2-k^2-i\epsilon$, $B=m^2 -(p-k)^2 -i\epsilon$,
$\epsilon \to 0^+$,  $M_\pi$ the pion mass and $d$ the number of space--time
dimensions. Any general
one--loop diagram with arbitrary many insertions from external sources
can be brought into this form by combining the propagators to a single
pion and a single nucleon propagator. It was also shown
that this procedure leads to a unique, i.e.
process--independent result, in accordance with the chiral Ward
identities of QCD \cite{BL}.  
Consequently, the transition from any one--loop graph $H$
to its IR singular piece $I$ defines a symmetry--preserving regularization.
For more details, the reader is referred to \cite{BL}.

\medskip\noindent
There is one special feature to IR regularized baryon CHPT that we have to discuss.
Here, we are interested in extending the range of photon virtualities $Q^2$ to 
values that eventually allow a matching to the perturbative QCD calculations. 
However, there is a generic limit to such an extension. A priori, infrared 
regularization defines a low energy effective
field theory, but due to the resummation of kinetic energy insertions and similar
effects, one might hope to extend this framework to higher virtualities than the
HBCHPT scheme (this expectation was borne out explicitely in the calculation of 
the neutron charge form factor, see \cite{KM}). Consider now a specific 
one--loop function in  $d$ dimensions 
\beq
\Gamma_0 (s,Q^2) = {1\over i} \, \int {d^d l \over (2\pi )^d} \, {1 \over
(M_\pi^2 -l^2)\,(m^2 - (p+q-l)^2)\,(m^2 -(p-l)^2) }~,
\eeq
with $(p+q)^2 = s$ and $p^2 = m^2$. In infrared regularization, 
this integral is written as
\beq\label{Ga0}
\Gamma_0 (s,Q^2) = \Gamma \left( 3 - \frac{d}{2}\right) \, \int_0^1 \, dy \,
\int_0^\infty \, dz \, z \, C^{\frac{d}{2} -3} (y,z)~,
\eeq
with $C$ a polynom in $y$ and $z$,
\beqa
C &=& A \,z^2 + B\, z +  \mu^2~, \nonumber \\
A &=& 1 + 2\,y\,\Omega\,\mu + y\, \mu^2 - y \, (1-y)\, \tau~, \nonumber \\
B &=& - 2\,y\,\Omega\,\mu - (1+y) \, \mu^2 ~,  
\eeqa
and
\beq
\Omega = {s -m^2 - M_\pi^2 \over 2 m M_\pi}~, \quad \mu = {M_\pi \over m}~, \quad 
\tau = {q^2 \over m^2}= -{Q^2 \over m^2}  ~.
\eeq
The $\Gamma$--function in Eq.(\ref{Ga0}) contains the pole as $d\to 4$ which has
to be dealt with in standard fashion. We also work with the standard value of the
scale of dimensional regularization setting $\lambda = m$.
All quantities considered here contain derivatives with respect to $s$ of such loop functions
at $s = m^2 +q^2 \leq m^2$ (see the discussion at the end of
Section~\ref{sec:V2CS}). At that point,
\beq
A = 1 + y\, \tau - y\,(1-y) \, \tau = 1 + y^2 \, \tau ~,
\eeq
so that $A = 0$ for $\tau = -1/y^2$. Consequently, $\Gamma_0 (s,Q^2)$ does not
converge for $|\tau| \ge 1$. Thus, the photon virtuality is bounded by the 
nucleon mass. In fact, for higher derivatives the influence of this singularity
is felt at smaller values of $Q^2$ so that we restrict the range of photon 
virtualities to $Q^2_{\rm max} = 0.3 \ldots 0.5\,$GeV$^2$ depending on the observable
one looks at.

\subsection{Structure functions}
Within this approach, we have calculated the reduced  structure
functions  $\bar{S}_{1,2}^{(p,n)} (0,Q^2)$,
generically called ${\cal S}$ in this section. 
The chiral expansion of ${\cal S}$ takes the form
\beq
\bar {\cal S} = \bar{\cal S}^{\rm tree} + \bar{\cal S}^{\rm loop}~.
\eeq
In our case, the tree level contribution stems from the remainder of the
Born graphs which lead to the following amplitudes (we omit writing down the
crossed term contributions)
\beq
T^{\rm Born} = {C(Q^2) \over s-m^2} + {\cal R}~,
\eeq
with $s = (p+q)^2$, $C(Q^2)$ can be expressed in terms of the
nucleon electromagnetic form factors and ${\cal R}$ denotes the non--pole (polynomial)
remainder from the Born diagrams. Only this latter contribution survives
the subtraction of the contribution from the elastic intermediate state (nucleon
pole term). Note also that there is no local contribution from ${\cal L}_{\pi N}^{(4)}$.
This can be understood easily. For constructing a gauge--invariant operator corresponding to
a two--photon observable, one needs two factors of the electromagnetic
field strength tensor $F_{\mu\nu}$, which leads to terms of fourth order. However,
such terms are always proportional to spin--independent terms like e.g. 
$\epsilon^\prime \cdot \epsilon$, i.e. they
correspond to spin--independent Compton scattering (see e.g. \cite{BKMcomp}).
To generate an operator with the additional factor containing the nucleon spin matrices,
one needs at least one extra power in small momenta, i.e. such terms can only
start at fifth order in the effective Lagrangian. 

\begin{figure}[H]
\centerline{
\epsfysize=5in
\epsffile{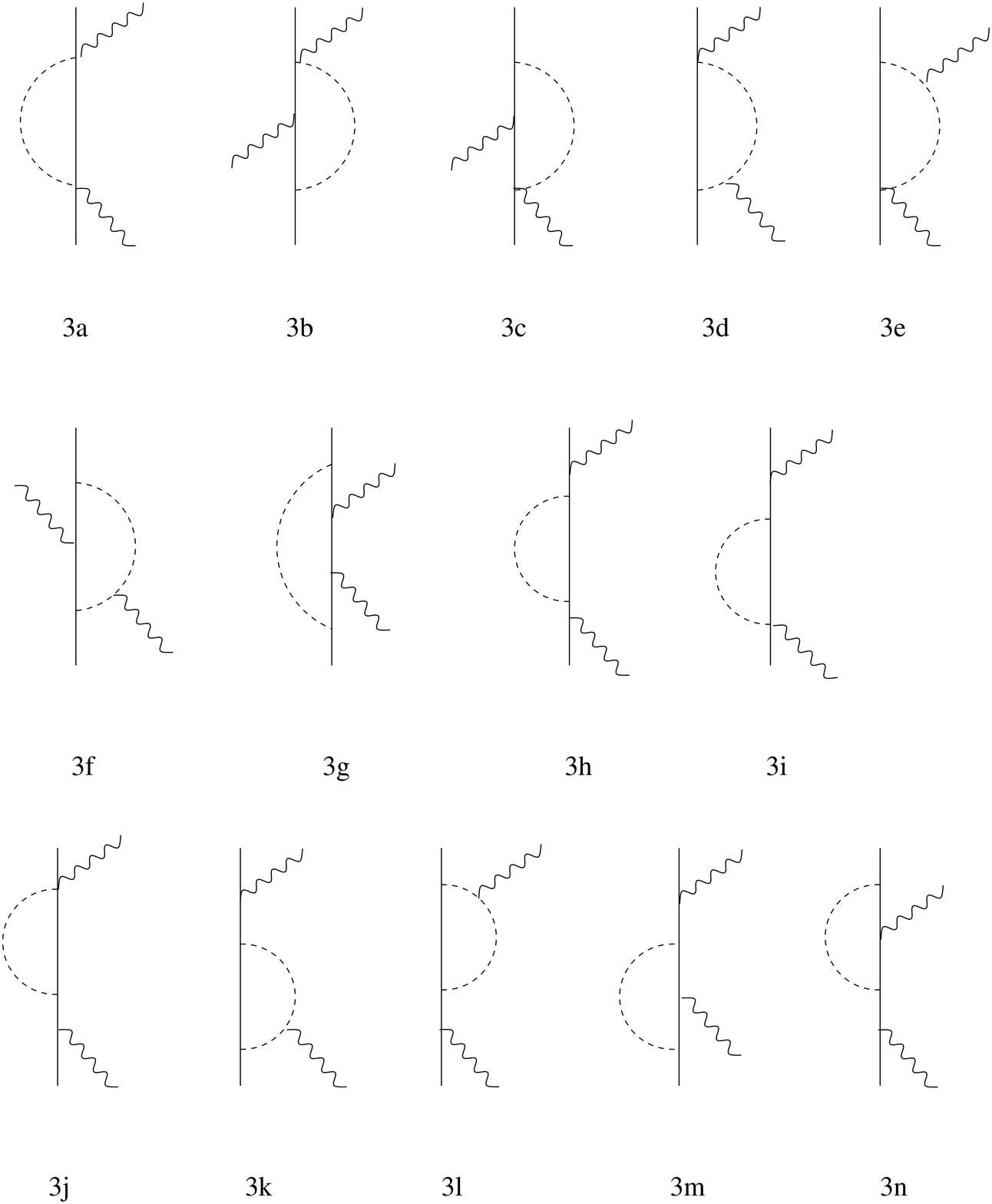}
}
\vspace{1cm}
\begin{center}
\caption{Feynman diagrams contributing at third order. Solid, dashed and wiggly
lines denote nucleons, pions and photons, in order. All insertions are from the
leading order (dimension one) effective Lagrangian. Crossed diagrams are not shown.
\label{fig:diag3}}
\end{center}
\end{figure}

\begin{figure}[H]
\vspace{3.5cm}
\centerline{
\epsfysize=5in
\epsffile{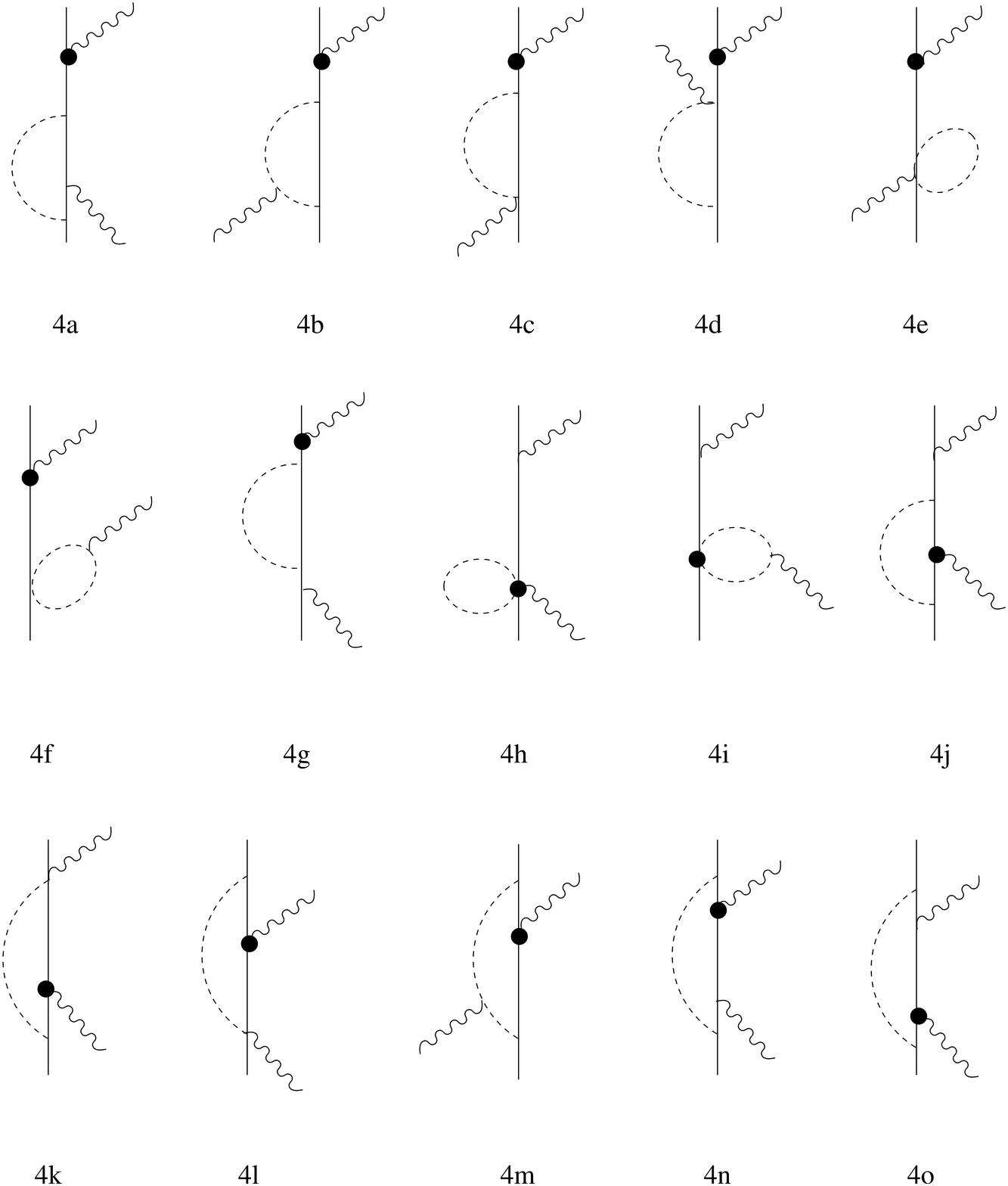}
}
\vspace{1cm}
\begin{center}
\caption{Topologically inequivalent Feynman diagrams contributing at fourth order. 
Solid, dashed and wiggly lines denote nucleons, pions and photons, in order. 
The black filled circle denotes an insertion from the dimension two effective 
Lagrangian. All other insertions are  from the leading order (dimension one) 
effective Lagrangian. Crossed diagrams are not shown.
\label{fig:diag4}}
\end{center}
\end{figure}

\medskip\noindent
The pertinent one--loop diagrams are shown in Figs.~\ref{fig:diag3} and \ref{fig:diag4}
for the third and   fourth order contributions, respectively. Note that different
to the heavy baryon approach, we do not need to consider 
dimension two insertions from the kinetic energy on the nucleon
lines since this is done automatically in a Lorentz--invariant
formulation. In fact, all such diagrams are counting as third order
in the approach used here, the only genuine fourth order graphs are
the ones with one insertion $\sim c_{6,7}$ (anomalous magnetic moment).
Note that these are indeed the only low--energy constants which enter in a complete
one loop calculation (whenever possible, we substitute these LECs by
the appropriate combinations of the proton and neutron magnetic moments).
Therefore, the  $\bar{S}_{1,2}(\nu,Q^2)$ are already non--vanishing at third order,
quite in contrast to the heavy baryon calculation.  As a check, we remark 
that in the limit of a very large nucleon mass, one recovers the 
earlier HBCHPT  results of \cite{BKMdhg}, \cite{Ji1} and also the
recent results of \cite{Kao}. However, as already found in the 
relativistic calculation of the slope of the generalized GDH 
sum rule in \cite{BKMdhg}, 
one can not give the results in closed analytical form. The explicit
expressions for the one--loop contributions to the structure functions
in terms of the pertinent loop functions are given in the appendix. 
Whenever possible, we will compare our results to the ones obtained
within HBCHPT. It should already be noted here that we find a very
strong dependence on the value of the nucleon mass, not untypical for
the nucleon spin sector.
Of course, there are also non--negligible  
resonance contributions to the $\bar{S}_i (\nu,Q^2)$. More precisely,
the effects from the $\Delta$ are expected to  largely cancel in $\bar{S}_1 (\nu,Q^2)$
\cite{Ji2} but are supposed to be more important in $\bar{S}_2 (\nu,Q^2)$. There are
additional (smaller) vector meson and higher baryon resonance
contributions. In the following section, we briefly discuss how to
estimate the contributions from these states.

\subsection{Modeling resonance contributions}
\label{sec:reso}

It is well-known that the excitation of the $\Delta (1232)$ plays a significant role 
in the spin sector of the nucleon. One  therefore would like to
include the delta as a dynamical degree of freedom in the effective
Lagrangian. At present, that can only be done systematically in the
heavy baryon scheme treating the nucleon--delta mass splitting as an
additional small parameter \cite{HHK}. An effective field theory formulation for
the relativistic  pion--nucleon--delta system does not yet exist. Therefore,
to get an estimate of the contribution of the $\Delta$-resonance to 
the various spin structure functions discussed here,
we calculate relativistic Born graphs. 
These are obtained using the well--known relativistic
spin-3/2 propagator (with momentum $P$, running from $\nu$ to $\mu$):
\beqa
-i {\barre P +m_\Delta \over P^2 -m_\Delta^2} \biggl(g_{\mu\nu}
-{1\over3}\gamma^\mu \gamma^\nu -{2 P_\mu P_\nu \over 3 m_\Delta^2}
 +{P_\mu \gamma^\nu-P_\nu \gamma^\mu \over 3 m_\Delta}\biggr)~,
\eeqa
and the  $\Delta \to N \gamma$ transition operator (for an in--coming
photon with momentum $k$ and an out--going  nucleon with  
momentum $p$):
\beqa
&-&{ie g_1 \over 2m} \gamma_5
\biggl(\epsilon^\mu \barre k   -k^\mu \barre \epsilon +
(Y+{1\over2})(\barre \epsilon \barre k- \barre k \barre \epsilon)\gamma^\mu
\biggr)T^3 \\
&+&{ie g_2 \over 4m^2}\biggl[ p \cdot k \gamma_5
\biggl(\epsilon^\mu-(X+{1\over2}) \barre \epsilon\gamma^\mu
\biggr)-\epsilon \cdot p\biggl(k^\mu-(X+{1\over2})\barre k\gamma^\mu
\biggr)\biggr]T^3~.
\eeqa
Here, $T$ is the ${1 \over 2} \to {3 \over 2}$ isospin transition operator and
$m_\Delta$ is the delta mass.
The $\Delta \to N \gamma$ transition depends on two off-shell parameters 
$X$, $Y$ and two transition
strengths $g_1$ and $g_2$, quantities which are not so well
known. We stress that in an effective field theory approach such a dependence
on off--shell parameters would be lumped in certain low--energy constants, however,
in the absence of such a systematic analysis of delta effects we must resort to the
procedure adopted here.  Bounds on $X, Y, g_1$ and $g_2$
have been given in Ref.~\cite{BKMcomp}:~$-0.8 \le X,Y \le 0.4, 4 \le g_1 \le 5$ and
$4.5 \le g_2 \le 9.5$. Here we constrain $g_1$ to its large $N_c$ relation,   
$g_1=3 (1+\kappa_p-\kappa_n)/ 2 \sqrt{2} =5.0$ and 
use two sets of parameters, $X=Y=0.4$, $g_2=4.5$, and $X=Y=-0.8$, $g_2=9.5$, 
respectively.
We note that these bounds are very conservative, a more precise determination based
on a combined reanalysis of spin--independent Compton scattering and pion electroproduction
based on IR baryon CHPT would certainly lead to more stringent bounds. Furthermore, 
in order to take into account the fact that the  $\Delta \to N \gamma$
transition occurs at finite $Q^2$ we entertain the possibility of introducing a 
transition form factor $G_{\Delta N \gamma} (Q^2)$, following \cite{EKPW}
\beqa\label{ff}
G_{\Delta N \gamma} (Q^2) = 
{{\rm exp}(-0.2 \, Q^2/{\rm GeV}^2) \over (1+1.41 \, Q^2/{\rm GeV}^2)^2}
\eeqa
as extracted from pion electroproduction in the $\Delta(1232)$-resonance
region, see Refs.~\cite{trans}. Note, however, that the energies and virtualities
of these data were sizeably larger than the range of values considered here. In a way,
such a form factor subsumes some higher order effects, it any way only plays a significant
role for virtualities $Q^2 > 0.15\,$GeV$^2$. 
We will come back to this issue when we
discuss the result in the next section.
Of course, there are also smaller contributions from higher baryon
resonances, but we do not include them in this work.

\medskip\noindent
A less pronounced though important resonance contribution is related
to the vector mesons. Again, a systematic EFT prescription how to 
include these degrees of freedom does not exist. We adopt here the
procedure advocated in Ref.\cite{KM}.
In the pion--nucleon EFT, any vector meson 
contribution is hidden in the values of the various LECs. However,
the momentum dependence of the vector meson propagator is only
build up slowly by adding terms of ever increasing chiral dimension.
This can be done much more efficiently by including vector mesons
in a chirally symmetric manner and retaining the corresponding 
dimension two counterterms, so that the LECs $c_{6}, c_7$ are effectively
replaced by \cite{KM}
\beqa
c_6 &\to& \hat{c}_6 
+ g_{\rho NN} \, \kappa_\rho \, \frac{F_\rho M_\rho}{M_\rho^2-t} ~, \nonumber
 \\
c_7 &\to& \hat{c}_7 
- \frac{g_{\rho NN} \, \kappa_\rho}{2} \, \frac{F_\rho M_\rho}{M_\rho^2-t}
+ \frac{g_{\omega NN} \, \kappa_\omega}{2} \, \frac{F_\omega M_\omega}{M_\omega^
2-t}
+ \frac{g_{\phi NN} \, \kappa_\phi}{2} \, \frac{F_\phi
  M_\phi}{M_\phi^2-t} ~. 
\label{c67VM}
\eeqa
Here, $t$ is the invariant four--momentum squared and the remainders 
$\hat{c}_6,\hat{c}_7$ account for physics not
related to vector mesons. They have been determined from fitting the
nucleons electromagnetic radii \cite{KM}. All other parameters
appearing in Eqs.(\ref{c67VM}) can be taken form the
dispersion--theoretical analysis of Refs.~\cite{MMD}. The values of these
parameters as used here are $M_\rho = 770\,$MeV,  $F_\rho = 152.5\,$MeV,
$g_{\rho NN} = 4.0$, $\kappa_\rho = 6.1$, $ M_\omega = 780\,$MeV,
 $F_\omega = 45.7\,$MeV,
$g_{\omega NN}= 41.8\,$, $\kappa_\omega = -0.16$, 
 $ M_\phi = 1020\,$MeV, $F_\phi = 79.0\,$MeV,
$g_{\phi NN}= -18.3$ and $\kappa_\phi = -0.22$.

\section{Results and discussion}
\def\theequation{\arabic{section}.\arabic{equation}}
\setcounter{equation}{0}
\label{sec:res}

Before presenting results, we must fix parameters.
We use $g_A = 1.267$, $F_\pi = 93\,$MeV, $M_\pi = 139.57\,$MeV, $m =
938.9\,$MeV, $\kappa_v =3.706$ and $\kappa_s = -0.120$. The latter
two quantities include the LECs $c_6$ and $c_7$. The pertinent
delta and vector meson parameters were already given in the preceding
section. We stress again that no unknown parameters enter our calculation
so all the following results are true predictions to fourth order (more
precisely, this refers to the chiral expansion of the various quantities).

\subsection{Complete fourth order results} 
\label{sec:four}
\medskip\noindent
We first discuss the two structure functions, the 
various sum rules and the moments thereof.
Results on the chiral expansion for $\bar S_1^{p,n} (0, Q^2)$ and 
$\Gamma_1^{p,n} (Q^2)$ as well as
for $\Gamma_1^{p-n} (Q^2)$ were already given in \cite{BHMdhg}, together with
a comparison to the HBCHPT results. Here, we show in the Fig.~\ref{fig:s1pn} 
the one--loop predictions for $\bar S_1^{p,n} (0, Q^2)$
supplemented by the vector meson and delta contributions
calculated as described in Section~\ref{sec:reso}. The band associated to  the resonance
contributions is due to the parameter variations of certain delta couplings as discussed
before. Of course, to draw a final conclusion on the size of the resonance part one
has to include $\pi \Delta$ loops in a systematic manner (for some attempts see
\cite{BKMdhg,Ji2,Kao}). No published data exist so far
for a direct comparison with these predictions, for orientation we show the
prediction for $\Gamma_1^p (Q^2)$ including resonance effects in 
Fig.~\ref{fig:gamma1p}.\footnote{We remark that in some terms of higher order we did not
replace the combinations of the LECs $c_6$ and $c_7$ by $\kappa_p$ and $\kappa_n$ 
in  \cite{BHMdhg} which
leads to slightly different results for the structure functions.}
In Figs.~\ref{fig:s12pn} and \ref{fig:s2pn} we  show the one--loop
results for the first moment $\bar S_1^{(2)} (Q^2)$
as defined in Section~\ref{sec:sumrules} and
for $(Q^2/m\nu) \, \bar S_2 (\nu,Q^2)|_{\nu = 0}$ (this is the quantity
which indeed enters the various sum rules)  for the proton and the neutron.
The results are similar to the ones presented for $\bar S_1^{p,n} (0, Q^2)$
in \cite{BHMdhg}. The fourth order corrections are not overly large for
virtualities below $Q^2 \simeq 0.15\,$GeV$^2$ and are very different
from the ones obtained in HBCHPT. These differences can be understood from the
very strong mass dependence of these quantities, the heavy baryon result can
be obtained for large nucleon masses, but the loop corrections change rather
dramatically as the mass is varied. We mention in particular the complete resummation
of the kinetic energy insertions in our approach, whereas these are only build
up order by order in the heavy fermion scheme. Stated differently, the $1/m$
expansion underlying HBCHPT is only slowly converging for the spin--dependent
structure functions considered here.
The delta Born and vector meson contributions to 
$(Q^2/m\nu) \, \bar S_2 (\nu,Q^2)|_{\nu = 0}$
are displayed in Fig.~\ref{fig:s2pnr}. First, these contributions lead to a visible
change in particular for the proton and second, the uncertainty is smaller as for
$\bar S_1^{p,n} (0,Q^2)$ chiefly because the second structure function does not
depend on the parameter $Y$ \cite{EKPW}. Again, the inclusion of $\Delta \pi$ loops
is called for.

\begin{figure}[H]
\vspace{0.9cm}
\parbox{.49\textwidth}{\epsfig{file= 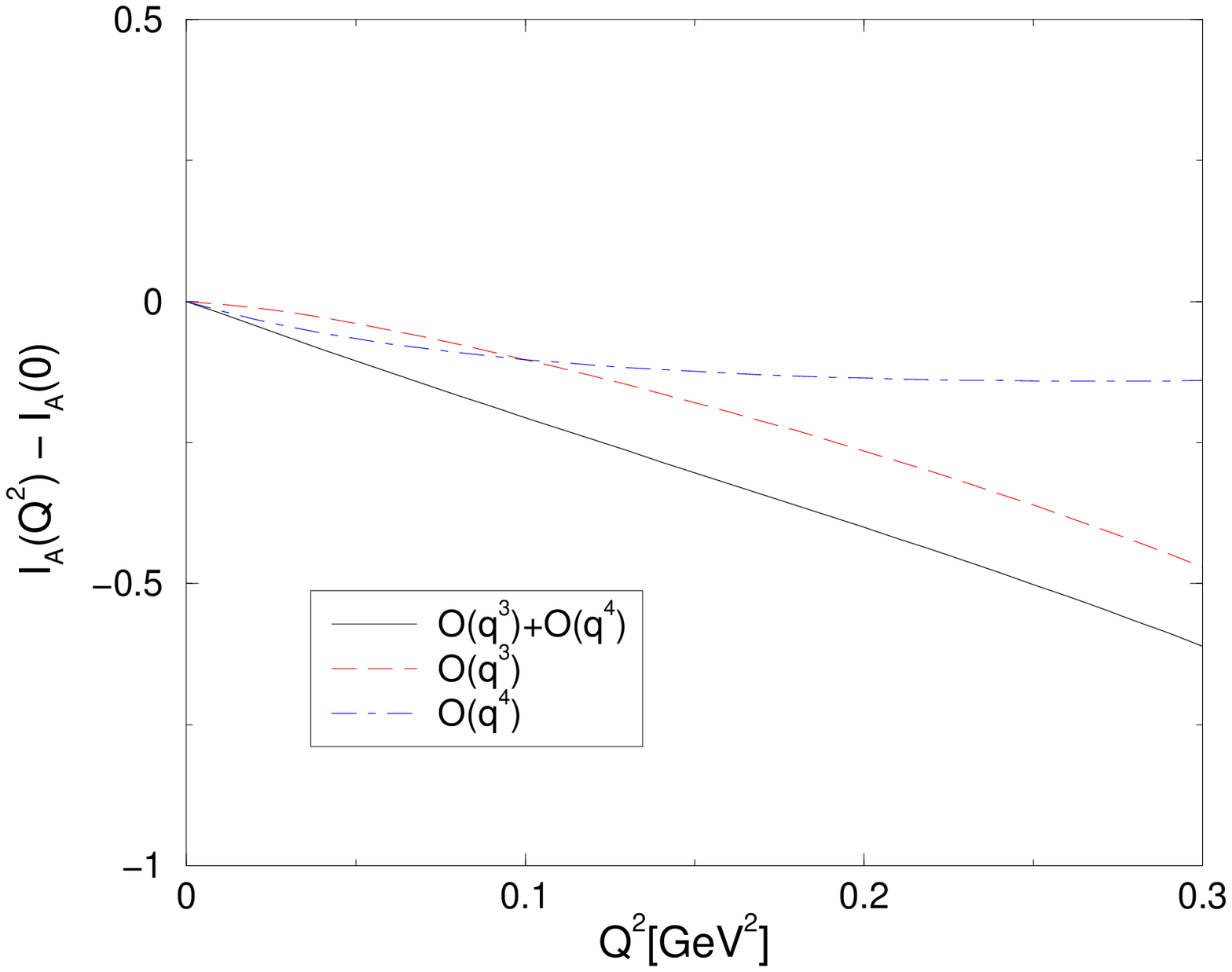,width=.48\textwidth,silent=,clip=}}
\hfill
\parbox{.49\textwidth}{\epsfig{file= 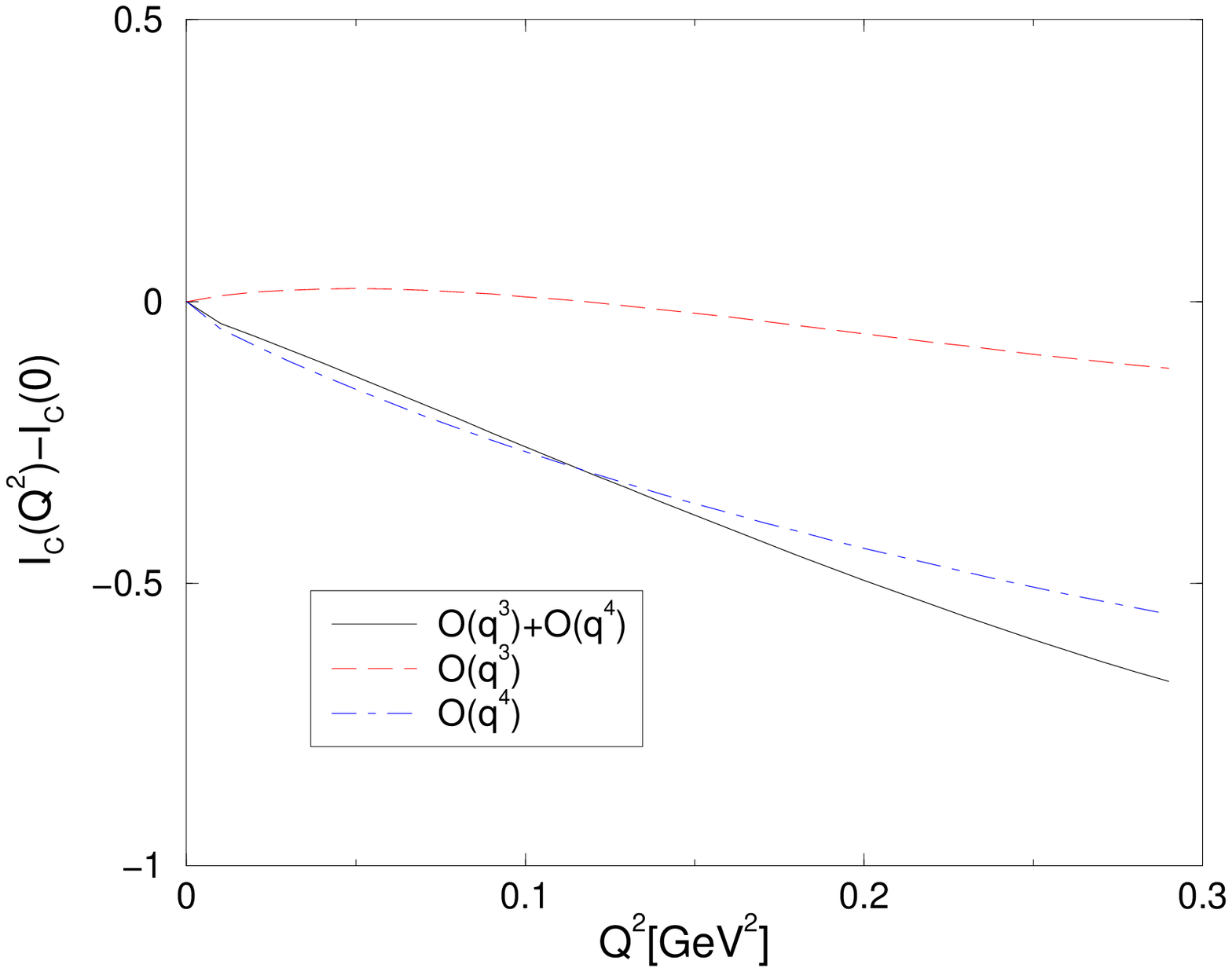,width=.48\textwidth,silent=,clip=}}
\vspace{0.3cm}
\begin{center}
\caption{The integrals $I_A (Q^2) - I_A (0)$ (left panel) and 
   $I_C (Q^2) - I_C (0)$ (right panel) for the neutron in units of GeV$^{-2}$.  
   The solid line gives the fourth order result, the third and fourth order 
   contributions are depicted by the dashed and dot--dashed lines, respectively.
\label{fig:IAC}}
\end{center}
\end{figure}

\medskip\noindent
Recently, data from Jefferson Lab on the momentum dependence 
of the generalized DHG sum rule for the neutron at low and intermediate $Q^2$
have become available \cite{E94}. In Fig.~\ref{fig:IAndat}
we show the lowest two data points for the integral $I_A^n (Q^2)$ in comparison
to our theoretical prediction. Consider first the left panel, which refers to the
results without the form factor of Eq.(\ref{ff}). As stressed before, the theoretical
uncertainty is rather large due to the large range of values of delta parameters.
It is, however, remarkable that the trend of the data is well described for
values of $g_2, X, Y$ at the upper edge of their allowed values (in magnitude).
Positive values for $X,Y$ and the smaller values of $g_2$ are clearly excluded
from this comparison. For illustration, we show that with the set of parameters
$g_2 =9.5$, $X = Y = -0.725$ one obtains a good description of the data (see the
dot--dashed line).
In the right panel of that figure, we show the results if
in addition the form factor is included. As expected, the allowed band shrinks and
furthermore, for the second set of delta parameters the theoretical
description turns out to be 
in rather good  agreement with experiment for $Q^2 \lesssim 0.15$~GeV$^2$. 
The bending of the data to lower values with increasing virtuality
cannot be explained any more, which presumably points towards the inadequacy of the
form factor Eq.(\ref{ff}) for analysing the new precision data.\footnote{We remark
that the theoretical band based on our preliminary calculation shown in \cite{E94}
is somewhat broader because we allowed for the delta coupling $g_2 = X = Y = 0$
which we now consider excluded.}
%
It is also interesting  to compare the chiral loop prediction for the
integrals $I_A (Q^2)$ and $I_C (Q^2)$, as shown in Fig.~\ref{fig:IAC} for the neutron.
While the third and fourth order contributions are rather different, the total one--loop
prediction for $I_A^n (Q^2)$ and $I_C^n (Q^2)$ comes out almost equal, as expected,
see Ref.~\cite{Drechsel}. 

\begin{figure}[H]
\vspace{1.1cm}
\parbox{.49\textwidth}{\epsfig{file= 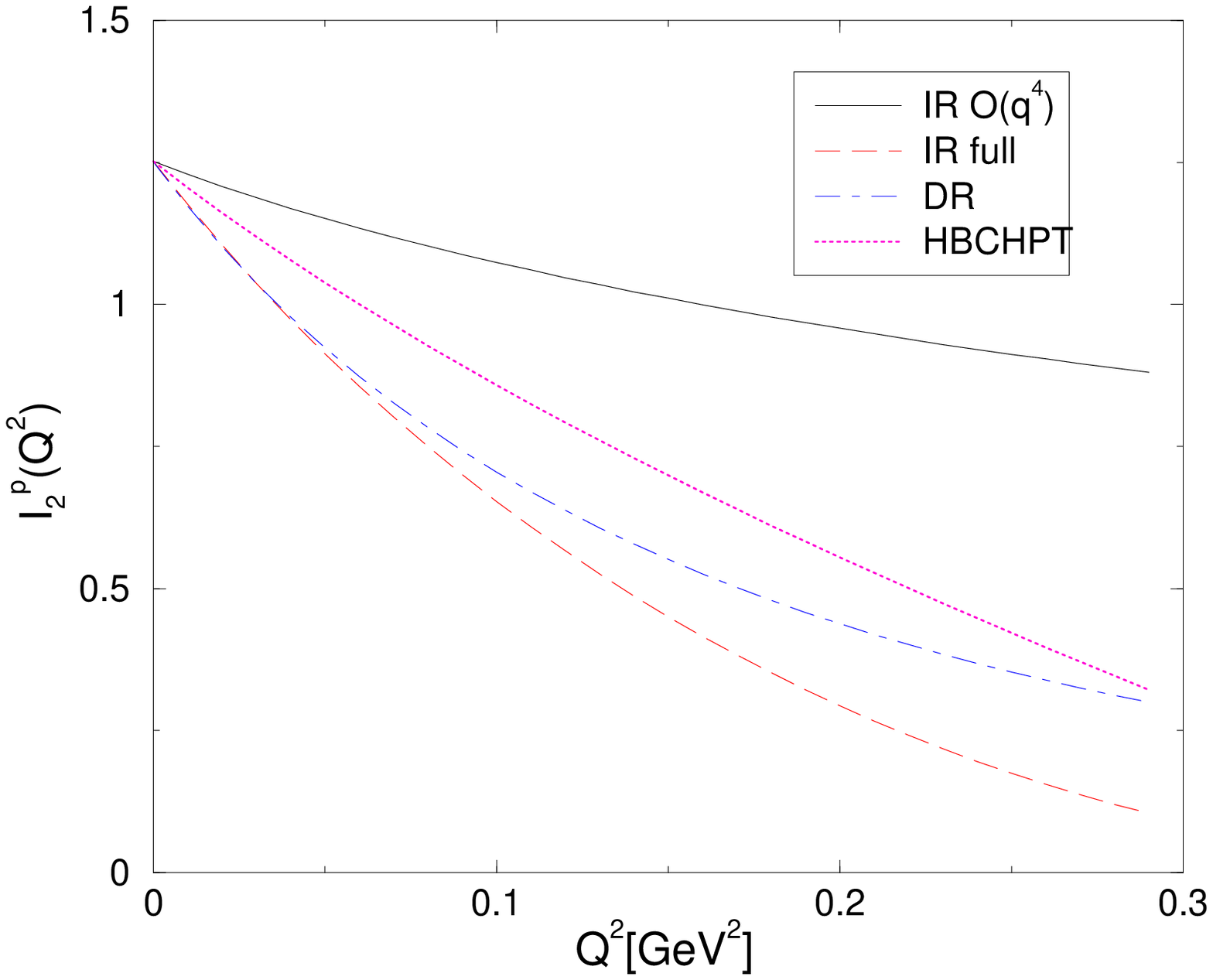,width=.48\textwidth,silent=,clip=}}
\hfill
\parbox{.49\textwidth}{\epsfig{file= 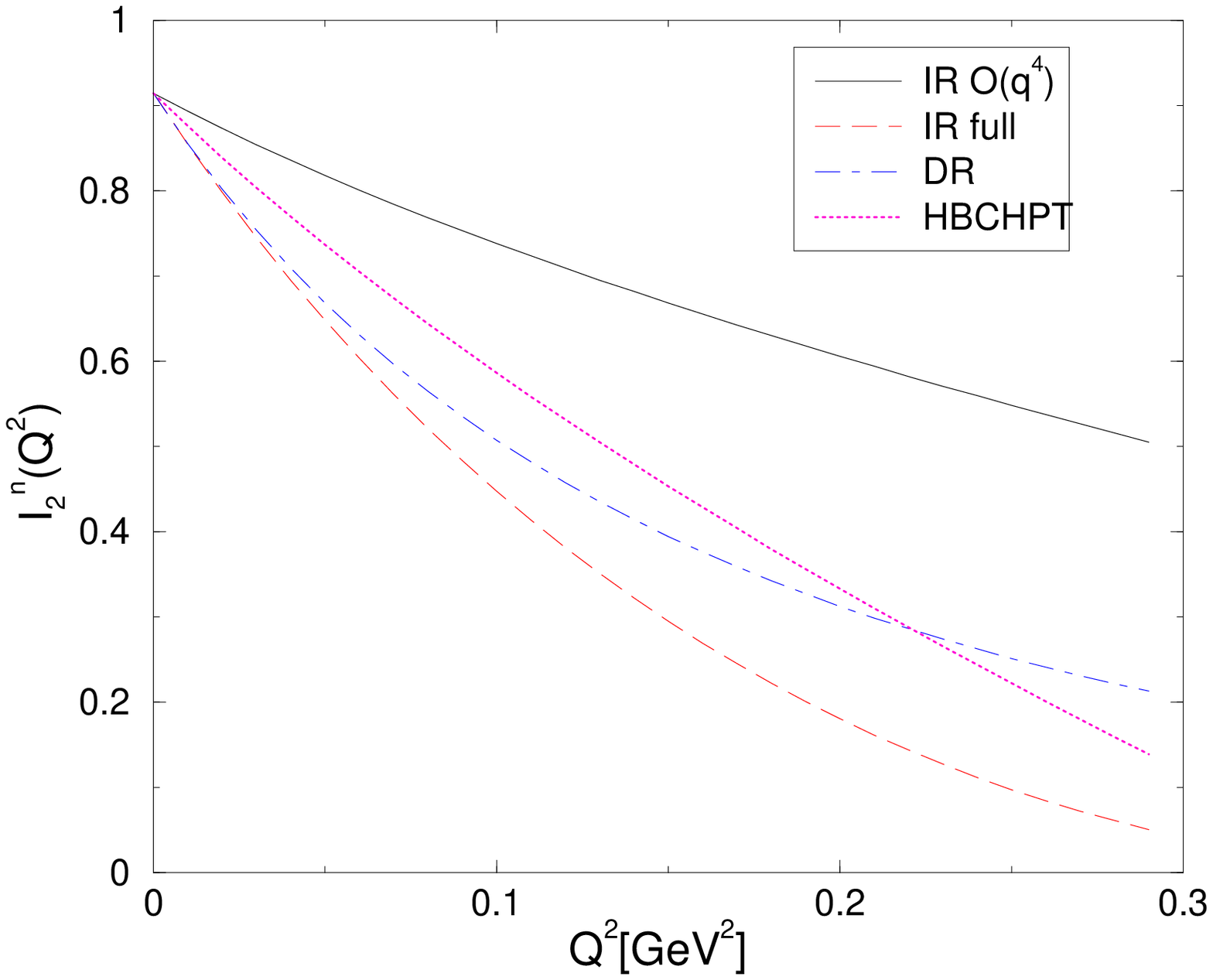,width=.48\textwidth,silent=,clip=}}
\vspace{0.3cm}
\begin{center}
\caption{\label{fig:I2pn}
          The integral $I_2 (Q^2)$ for the proton (left panel) and the neutron 
          (right panel). The solid line gives the complete fourth
          order result, that is the product of the form factors is
          taken to that order. Dashed (dot-dashed) line: taking into account the full fourth
          order (dispersion-theoretical) result for the form factors
          (i.e. without truncating the product of the form factors). 
          The HBCHPT result of \protect \cite{Kao} is also shown (dotted line).
}
\end{center}
\end{figure}

\medskip\noindent
We now turn to the BC integrals $I_2 (Q^2)$ for the
proton and the neutron as shown in Fig.~\ref{fig:I2pn}. First, we note
that the validity of the BC sum rule was shown for HBCHPT in
\cite{Kao}, it is also fulfilled if one uses infrared regularization (which is
simply a consequence of some general principles like e.g. crossing symmetry).
In  Fig.~\ref{fig:I2pn} we show the result for $I_2$ from the IR
calculation of \cite{KM}  where the form factors
are taken up to ${\cal O}(q^4)$ (corresponding to the curve denoted as IR:~full). 
The difference with the
dispersion--theoretical result \cite{MMD} is mostly accounted for by 
vector mesons and one therefore obtains a rather good description of $I_2$
in IR plus vector mesons.  However, strictly speaking  this does not
correspond to an  ${\cal O}(q^4)$ calculation of $I_2$, in which one
only retains terms in the product of the form factors
up-to-and-including fourth order (corresponding to the curve denoted as 
IR:~{\cal O}($q^4$)). It turns out that this ${\cal O}(q^4)$ calculation is rather
far from the empirical results since higher order terms are important.
We come back to this issue in  Section~\ref{sec:five}.
In fact, it turns out that the HBCHPT result of Ref.~ \cite{Kao}  looks better 
though individually the form factors are described slightly 
worse.\footnote{Note that one has to be careful when comparing the large mass
limit from the IR calculation with the HBCHPT results. For example, the LECs
used in \cite{KM} do not correspond to the ones used in the HBCHPT calculation
of \cite{BFHM} and thus a seemingly contradicting result emerges.} However,
in that paper the Dirac form factor is simply a constant (that is the charge
of the nucleon) and the complete $Q^2$--dependence is thus given by the
Pauli form factor, quite in contrast to our calculation and the empirical
result. Note also that the HBCHPT result shows no curvature in contrast to 
the empirical curve. Given these observations, we do not share the opinion
of the authors of \cite{Kao} that the BC sum rule is a good
observable to bridge the gap between low and high $Q^2$.

\begin{figure}[H]
\parbox{.49\textwidth}{\epsfig{file= 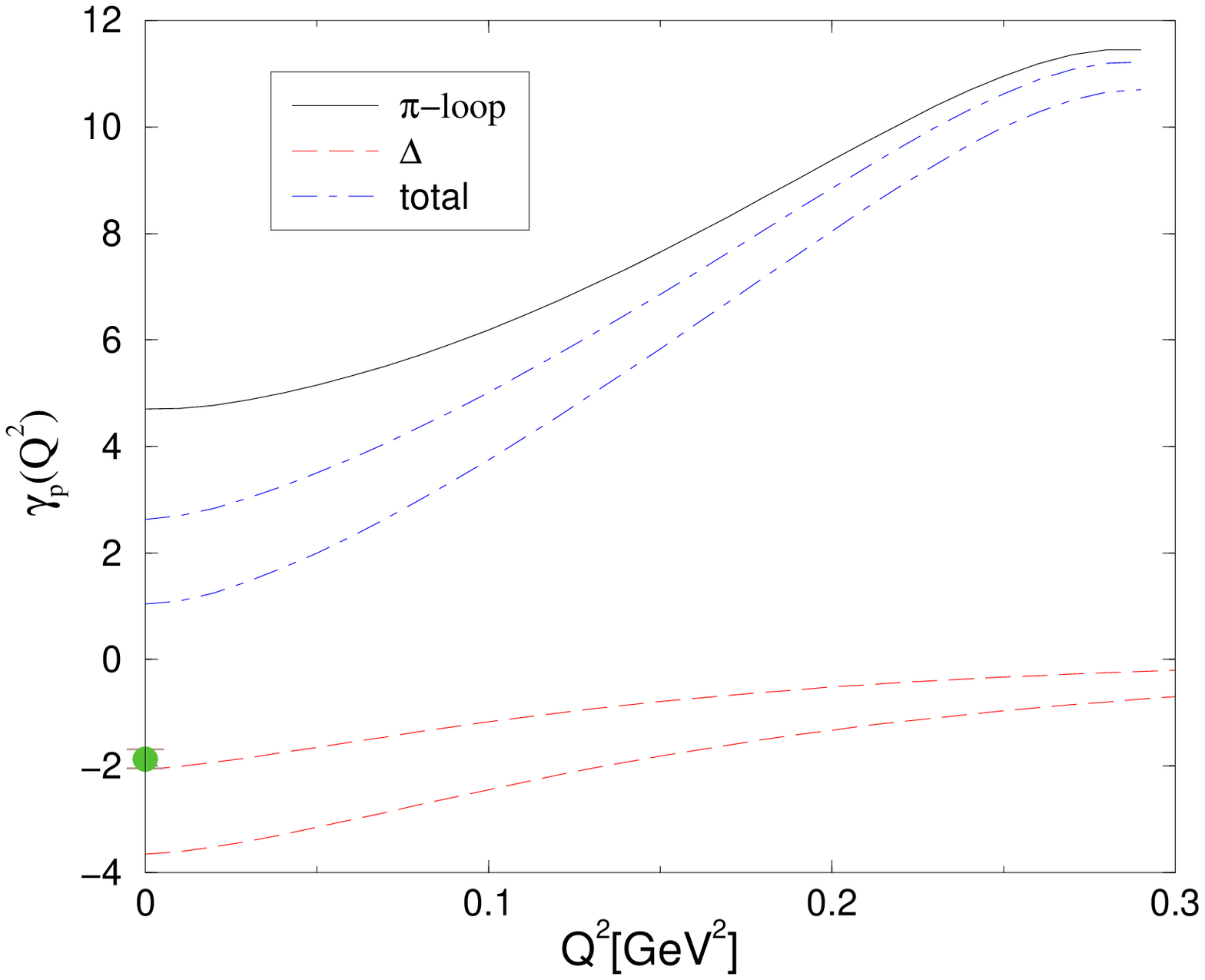,width=.48\textwidth,silent=,clip=}}
\hfill
\parbox{.49\textwidth}{\epsfig{file= 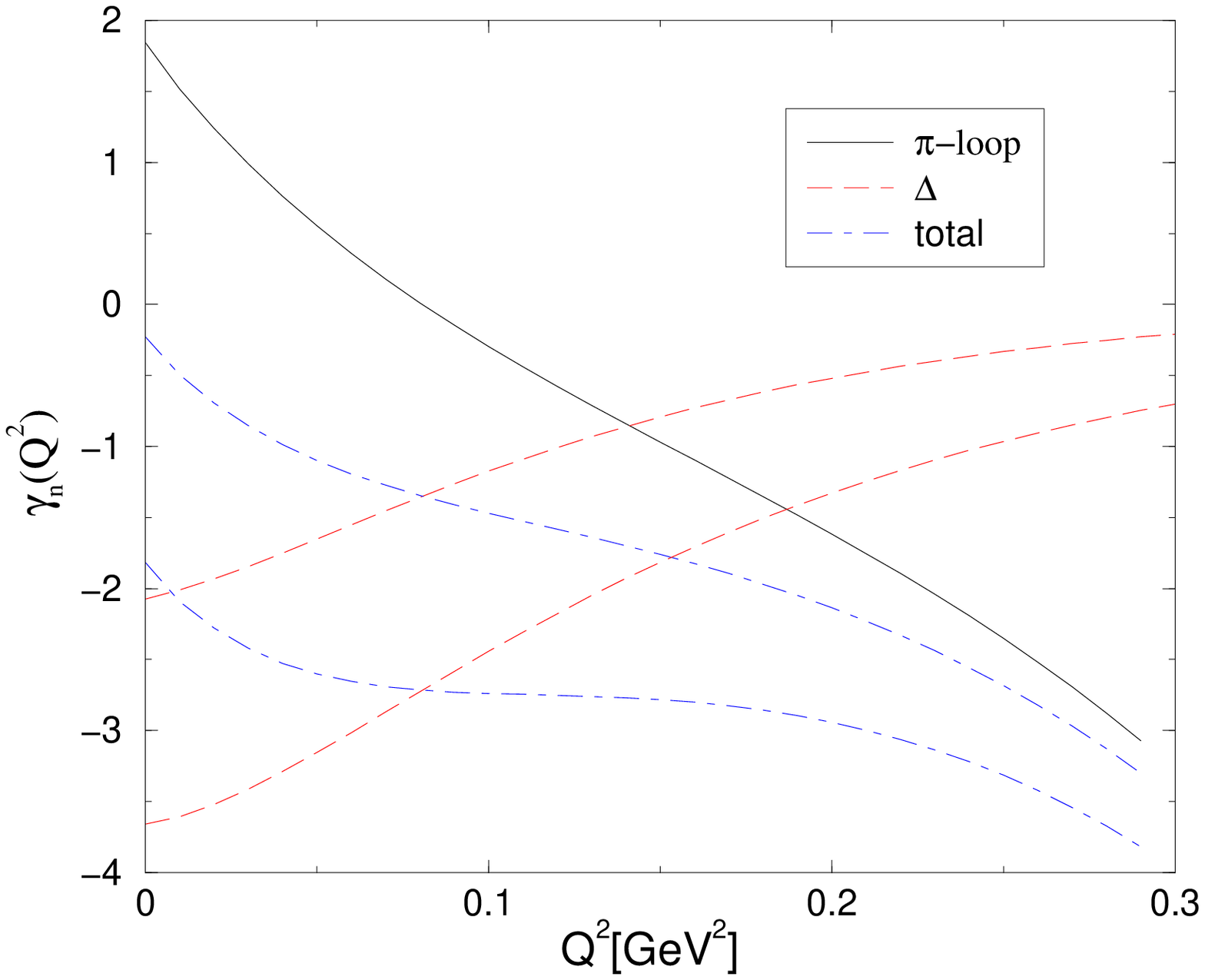,width=.48\textwidth,silent=,clip=}}
\vspace{0.9cm}
\begin{center}
\caption{\label{fig:gammapn}
          The forward spin polarizability $\gamma_0$ at finite virtuality for the proton 
          (left panel) and the neutron (right panel) in units of $10^{-4}$~fm$^4$. 
          The solid line gives the
          fourth order pion loop result, the delta contribution is given by the
          band spanned by the dashed lines. The total prediction is depicted by
          the dot--dashed lines. The MAMI data point on $\gamma_0^p
          (Q^2 =0)$ is shown by the circle \protect\cite{Ahrens}.
}
\end{center}
\end{figure}

\medskip\noindent
In Fig.~\ref{fig:gammapn}, we show the forward spin polarizability
for the proton and the neutron at finite photon virtuality. For the
proton, the chiral loop contribution is positive and increasing with
increasing $Q^2$, while the delta Born graph contribution is negative
and essentially flat. In total, the sum is positive, at odds with
the measured value of the spin polarizability (at the photon point)
\cite{Ahrens}. At this point we can only speculate that a combination of higher 
order and  delta loop effects will account for the discrepancy. This deserves further
investigation (see also the discussion in Section~\ref{sec:five}). 
The chiral expansion of $\gamma_0^p$ will be discussed 
below. For the neutron, the results are markedly different. While the
chiral loop contribution decreases with increasing photon virtuality,
the delta Born contribution rises, leading to a slightly decreasing sum.
Note also that $\gamma_0^n$ is negative in contrast to the positive $\gamma_0^p$,
our theoretical bounds are:
\beqa\label{gammapn}
&& \phantom{-}1.0  \cdot 10^{-4}\,{\rm fm}^4 
\le \gamma_0^p \le \phantom{-}2.6  \cdot 10^{-4}\,{\rm fm}^4~,
\\
&& -1.8 \cdot 10^{-4}\,{\rm fm}^4 \le \gamma_0^n \le -0.2 \cdot 10^{-4}\,{\rm fm}^4~.
\eeqa
One should not be worried about the difference of these numbers to the results
presented in \cite{EKPW} because in that paper a third order dimensionally
regularized relativistic calculation
was performed, which has no consistent underlying power counting.
It is interesting to consider the chiral expansion of pion loop
contribution to the forward spin polarizability of the proton and the
neutron.
These can be given in closed analytical form,\footnote{Note that the
terms of order $\mu^0$ (and higher) will be modified by two--loop
corrections (but not the terms $\sim \mu^{-2}$ and $\sim \mu^{-1}$).}
\beqa
\gamma_0^p&=&{e^2 g_A^2 \over 96 \pi^3 M_\pi^2 F_\pi^2}
\Biggl(1 -\biggl[{1 \over 8} \left( 21 + 3 \kappa_v+\kappa_s\right)\,
\pi\, \mu
+ \biggl( \bigl( 20 + {9 \over 2}(\kappa_v + \kappa_s) \bigr) \, \mu^2 
\nonumber \\
& &+ \, \bigl(26 + 7\kappa_v + 7\kappa_s) \, \mu^2 \ln\mu  \biggr) 
-{15 \over 64} \bigl(125+49 \kappa_v+63\kappa_s \bigr) \, \pi\, \mu^3
\biggr]  + {\cal O}(\mu^4) \, \Biggr) \nonumber \\
&=& \underbrace{4.45}_{{\cal O}(\mu^{-2})} 
- \underbrace{8.31}_{{\cal O}(\mu^{-1})}   
+ \underbrace{6.03}_{{\cal O}(\mu^0)}  
+ \underbrace{3.22}_{{\cal O}(\mu^1)}  
+ {\cal O}(\mu^2) \nonumber \\
&=& 4.64~, \\
\gamma_0^n &=& {e^2 g_A^2 \over 96 \pi^3 M_\pi^2 F_\pi^2}
\Biggl(1 -\biggl[{1 \over 8} \left( 9 + 3 \kappa_v-\kappa_s\right)\,
\pi\, \mu
+ \biggl( \bigl( 5 + 3\kappa_v - 3\kappa_s \bigr) \, \mu^2 
\nonumber \\
& &+4 \, \bigl(1 +\kappa_v -\kappa_s) \, \mu^2 \ln\mu  \biggr) 
-{15 \over 64} \bigl(5+15 \kappa_v-21\kappa_s \bigr) \, \pi\, \mu^3
\biggr]  + {\cal O}(\mu^4) \, \Biggr) \nonumber \\
&=& \underbrace{4.45}_{{\cal O}(\mu^{-2})} 
- \underbrace{5.25}_{{\cal O}(\mu^{-1})}   
+ \underbrace{2.00}_{{\cal O}(\mu^0)}  
+ \underbrace{0.68}_{{\cal O}(\mu^1)}  
+ {\cal O}(\mu^2) \nonumber \\
&=& 1.82~,
\eeqa     
where the last number refers to the full (not expanded) result and
all numbers are given in the canonical units of $10^{-4}\,$fm$^4$.
The first term in these series agrees, of course, with the
result of \cite{BKKM}. More interesting are, however, the higher
order terms which have stirred some controversy within HBCHPT
calculations, see \cite{Kumar,Ji3,GHM,Birse,GHMr} (more precisely,
this refers to the terms $\sim \mu^{-1}$). 
Let us consider first the proton. Upon
inspection, the chiral expansion does not seem to converge.
However,  we note  that the
sum of the four first terms in the expansion is 
$5.39 \cdot 10^{-4}\,$fm$^4$, the difference with the full result coming
essentially from the ${\cal O}(\mu^2)$ term, which brings another 
$-0.64 \cdot 10^{-4}\,$fm$^4$. Naively, one would expect that the
expansion parameter is $\mu \simeq 1/7$, but clearly the appearance
of large numerical prefactors, in particular due to the large value
of $\kappa_v \simeq 3.7$, invalidates such a simple picture. 
Things look somewhat different for the neutron. 
The sum of the first four terms amounts to $1.87 \cdot 10^{-4}\,$fm$^4$,
only a few percent off the full result. Note, however, the
almost complete cancellation between the first and the second term.
Therefore, to make a reliable prediction for the chiral loop
contributions to $\gamma_0^n$, one has to account for the terms
of order $\mu$ as done here. To get an idea about the theoretical
uncertainty for these numbers, we have replaced the physical values of
$\kappa_v$ and $\kappa_s$ in the above expressions by the values of the
LECs $c_6$ and $c_7$ to third order as given in \cite{KM}. To the accuracy
we are working, this is legitimate. The resulting values for the forward
spin polarizabilities are $\gamma_0^p = 4.95\cdot 10^{-4}\,$fm$^4$ and
$\gamma_n^0 = 1.70 \cdot 10^{-4}\,$fm$^4$, not very different from the
results given above. 
We have also identified the reason for the bad convergence, it is
entirely related to the contributions from the Born terms, as
suggested by the method used in \cite{GHM}. We thus give here
the third and fourth order pole and non--pole contributions to
$\gamma_0^{p,n}$ (note that to third order one has no pole contribution
in the case of the neutron):
\beq\label{propole}
\begin{array}{cccccccr}
\gamma_0^{p,3,{\rm pole}} & =& 0 & - \, 3.12 &+ \, 2.93 &+ \, 1.49 & -
\,0.29 + \dots 
&= \phantom{-}0.96~,  \\
\gamma_0^{p,3,{\rm non-pole} } & =& 4.45 &- \, 2.34 & - \, 0.02 & + \, 0.14 & + \ldots 
& =  \phantom{-}2.03~,   \\
\gamma_0^{p,4,{\rm pole}} & =& 0 & - \, 1.86 &+ \, 2.88 &+ \, 1.91 & -
\,0.41 + \dots 
&=   \phantom{-}2.45~,  \\
\gamma_0^{p,4,{\rm non-pole}} &= & 0 & - \, 0.93 & + \, 0  &- \, 0.12
& + \, 0.02 + \ldots
& = -0.81~,\\
\end{array}
\eeq
\beq\label{neupole}
\begin{array}{cccccccr}
\gamma_0^{n,3, {\rm non-pole}} & =& 4.45 &- \, 2.34 & + \, 0.36 & + \, 0.05 & + \ldots 
& =  \phantom{-}2.43~,   \\
\gamma_0^{n,4,{\rm pole}} & =& 0 & - \, 1.99 &+ \, 1.74 &+ \, 0.74 & -
\, 0.07 + \dots 
&=  \phantom{-}0.42~,  \\
\gamma_0^{n,4,{\rm non-pole}} &= & 0 & - \, 0.99 & + \, 0.24  &- \,
0.04 & - \, 0.02 + \ldots
& = -1.03~,\\
\end{array}
\eeq
where the first, second, third, $\ldots$ term refers to the terms of
order $1/M_\pi^2$, $1/M_\pi$, $M_\pi^0$, $\ldots$, respectively, and 
the last one is again the full (unexpanded) result at the given order.
All numbers are given in canonical units. One clearly sees that the pole
graphs are entirely responsible for the convergence problems, whereas
the non--pole terms show a good convergence at third as well as at fourth
order. One thus
might entertain the possibility of rearranging the power counting
accordingly. We come back to this point in Section~\ref{sec:five}.
Finally, note that a complete one--loop calculation to 
${\cal O}(q^4)$ within HBCHPT only gives the terms up--to--and--including 
${\cal O}(\mu^{-1})$. 

\begin{figure}[H]
\vspace{0.9cm}
\parbox{.49\textwidth}{\epsfig{file= 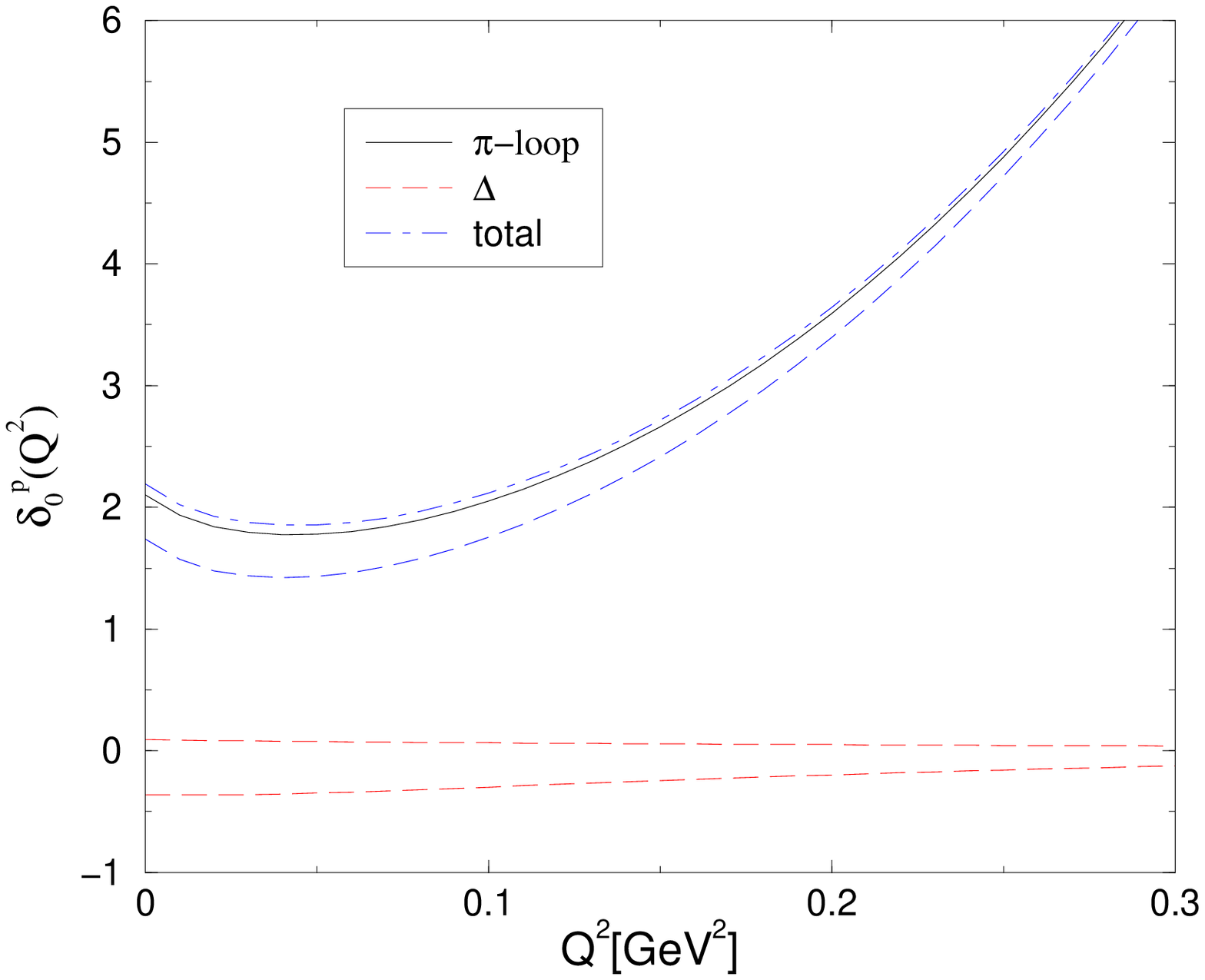,width=.48\textwidth,silent=,clip=}}
\hfill
\parbox{.49\textwidth}{\epsfig{file= 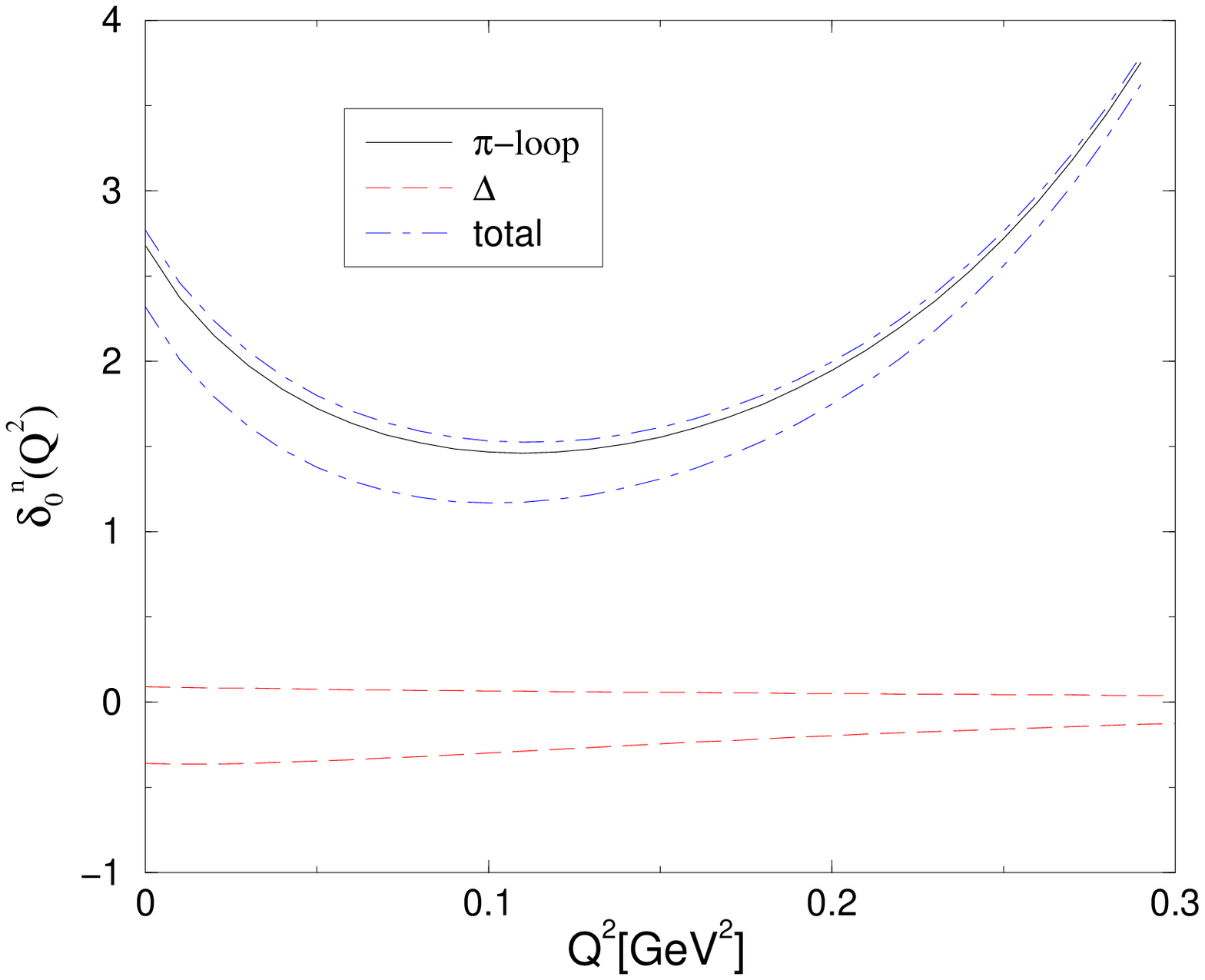,width=.48\textwidth,silent=,clip=}}
\vspace{0.3cm}
\begin{center}
\caption{\label{fig:deltapn}
          The longitudinal--transverse polarizability $\delta_0$  
          at finite virtuality for the proton 
          (left panel) and the neutron (right panel) in units of
          $10^{-4}$~fm$^4$. The solid line gives the
          fourth order pion loop result, the delta contribution is given by the
          band spanned by the dashed lines. The total prediction is depicted by
          the dot--dashed lines.
}
\end{center}
\end{figure}

\medskip\noindent
In Fig.~\ref{fig:deltapn} the longitudinal--transverse spin polarizability
for the proton and the neutron at finite virtuality is shown. Here, the
delta Born graph contribution is small, as can be easily understood making 
use of the multipole decomposition of Ref.~\cite{Drechsel}. Therefore, the
chiral prediction for $\delta_0^{p,n} (Q^2)$ is certainly more reliable
than for the forward spin polarizabilities. For $\delta_0^n (Q^2)$ as well as
for $\delta_0^p (Q^2)$  the slope at $Q^2 = 0$ is negative, consistent with the 
HBCHPT result of \cite{Kao}. At the photon point, we
have:
\beqa
&& 1.7  \cdot 10^{-4}\,{\rm fm}^4 \le \delta_0^p \le 2.2  \cdot 10^{-4}\,{\rm fm}^4~,
\\
&& 2.3 \cdot 10^{-4}\,{\rm fm}^4 \le \delta_0^n \le 2.8 \cdot 10^{-4}\,{\rm fm}^4~.
\eeqa
We note that these numbers are not very different from the leading $1/M_\pi^2$
singularity, which gives $\delta_0^p = \delta_0^n = 2.23 \cdot 10^{-4}\,$fm$^4$.
This can be made more transparent by considering the chiral expansion of these
quantities. We find
\beqa
\delta_0^p &=& {e^2 g_A^2 \over 192 \pi^3 M_\pi^2 F_\pi^2}
\Biggl(1  + \biggl[{1 \over 8} \bigl( -9 +  \kappa_v + 4\kappa_s\bigr)\,
\pi\, \mu
+ \biggl( \biggl(- 3 + \frac{3}{16}(15\kappa_v + 11\kappa_s) \biggr) \, \mu^2 
\nonumber \\
& &+ \, \frac{3}{8}\biggl(-\kappa_v +11\kappa_s \biggr)
 \, \mu^2 \ln\mu  \biggr) 
-{3 \over 64} \biggl( 119 - 15 (4 \kappa_v -5\kappa_s) \biggr) \, \pi\, \mu^3
\biggr]  + {\cal O}(\mu^4) \, \Biggr) \nonumber \\
&=& \underbrace{2.23}_{{\cal O}(\mu^{-2})} 
- \underbrace{0.75}_{{\cal O}(\mu^{-1})}   
+ \underbrace{0.53}_{{\cal O}(\mu^0)}  
+ \underbrace{0.12}_{{\cal O}(\mu^1)}  
+ {\cal O}(\mu^2) \nonumber \\
&=& 2.04~, \\
\delta_0^n &=& {e^2 g_A^2 \over 192 \pi^3 M_\pi^2 F_\pi^2}
\Biggl(1  + \biggl[{1 \over 8} \bigl( 3 +  \kappa_v - 4\kappa_s\bigr)\,
\pi\, \mu
+ \biggl( \biggl( 7 + 3\kappa_v - \frac{3}{2}\kappa_s \biggr) \, \mu^2 
\nonumber \\
& &+ \,12 \biggl(1  - \frac{1}{4}\kappa_s \biggr) \, \mu^2 \ln\mu  \biggr) 
-{3 \over 64} \biggl( 275 - 15 (\kappa_v + 4\kappa_s) \biggr) \, \pi\, \mu^3
\biggr]  + {\cal O}(\mu^4) \, \Biggr) \nonumber \\
&=& \underbrace{2.23}_{{\cal O}(\mu^{-2})} 
+ \underbrace{0.93}_{{\cal O}(\mu^{-1})}   
- \underbrace{0.26}_{{\cal O}(\mu^0)}  
- \underbrace{0.24}_{{\cal O}(\mu^1)}  
+ {\cal O}(\mu^2) \nonumber \\
&=& 2.66~,
\eeqa   
where again the number in the last row refers to the full one--loop result
and all numbers are given in units of $10^{-4}\,{\rm fm}^4$.
The chiral expansion of $\delta_0^p$ is very well behaved, furthermore, the
sum of the first four terms in the series is $2.13 \cdot 10^{-4}\,{\rm fm}^4$,
only 4\% off the full result.  For the neutron, we observe that the sum of the 
first four  terms, which is $2.66 \cdot 10^{-4}\,{\rm fm}^4$, essentially 
amounts to the total sum at one loop. The chiral expansion of $\delta_0^n$ is 
well behaved, although the terms of order $\mu^0$ are somewhat small.
We note that in contrast to the chiral expansion of $\gamma_0^{p,n}$
no  unnaturally large prefactors appear here. It thus appears that the
longitudinal--transverse spin polarizability is a better suited observable to
test the chiral dynamics of QCD. Finally, we have also calculated the 
longitudinal--transverse spin polarizabilities
using the third order values for $c_{6,7}$ instead of the physical values for
$\kappa_{s,v}$. This results in $\delta_0^p = 2.45 \cdot 10^{-4}\,{\rm fm}^4$
and $\delta_0^n = 3.13  \cdot 10^{-4}\,{\rm fm}^4$.

\subsection{Beyond fourth order} 
\label{sec:five}
As we have seen e.g. in the discussion of the forward spin polarizabilities, the
calculation of the Born graphs to fourth order is not quite sufficient in
some cases to obtain convergence. This observation was also behind the
modified method to obtain the spin polarizabilities from the
spin--structure amplitudes used in \cite{GHM}. We therefore assume that the missing 
physics largely resides in these Born terms, more specifically in the
leading two--loop corrections that appear at fifth order\footnote{For
  fifth order calculations within HBCHPT, see \cite{BMG}.}.
While a complete fifth order calculation
goes beyond the scope of this paper, it is straightforward to calculate the
contribution from the pole diagrams shown in
Fig.~\ref{fig:diag5} (for the combination of structure functions 
$S_1 (\nu,Q^2) - (Q^2/m \nu) \, S_2 (\nu,Q^2))$. 
This class of diagrams is gauge--invariant and,
of course, does not modify any LET or the DHG and BC sum rules. However,
new LECs related to dimension two insertions as in diagram 5t
and dimension four insertions as in diagram 5u appear.
These have, however, already been determined in the analysis 
of pion--nucleon scattering \cite{BM} and  of the
nucleons' electromagnetic form factors \cite{KM}. We use the same values,
i.e. $c_4 = 3.4\,$GeV$^{-1}$, $e_{54} = 0.26\,$GeV$^{-3}$ 
and $e_{74} = 1.65\,$GeV$^{-3}$~.
Other operators that appear in the chiral expansion of the nucleon
from factors \cite{KM} do not contribute to the particular combination
of the structure functions we are considering here.
Note that here we concentrate on the neutron, simply because there are
much less diagrams to calculate than in the case of the proton due to the
non--vanishing charge (dimension one insertion). For further calculational
simplicity, we restrict these considerations to observables that can be
derived from the combination of structure functions $S_1 (\nu,Q^2) - 
(Q^2/m \nu) \, S_2 (\nu,Q^2)$.
Of course, one should devise a modified power counting to properly
account for these effects. A resummation of these terms also to higher
order could be obtained by including the pertinent higher order
insertions in the baryon propagator. Whether this can be done in a
systematic fashion consistent with already existing successful calculations of
other observables remains to be seen. Here, we are only interested
to investigate whether these terms indeed contribute significantly in the 
problematic cases, these are the forward spin polarizability and  the
convergence of the BC sum rule.

\medskip\noindent
Consider first 
the BC sum rule for the neutron. Here, we have performed the complete
fifth order calculation since only Born terms contribute. As shown in 
Fig.~\ref{fig:I2n5}, there are sizeable corrections when going from fourth
to fifth order in the expansion of the product of the form factors. On the
other hand, for $Q^2 \lesssim 0.15\,$GeV$^2$, the fifth order result is not
very different from the ``full fourth order'' result (meaning that the form
factors are taken to fourth order, not their product).
Finally, we turn to the forward spin polarizability of the neutron. 
Using the same notation as in Eq.(\ref{neupole}), we find (in
canonical units):
\beq\label{pole5}
\begin{array}{cccccccccr}
\gamma_0^{n,5,{\rm pole}} & =& 0 & + \, 0 & + \, 0.02 &- \, 1.36 & - \,
0.18 & - \, 0.09 & 
+ \dots &= -1.41~,  \\
\end{array}
\eeq
where the first two zeros are consistent with the power counting, i.e. that
the two leading chiral singularities $\sim 1/M_\pi^2$ and  $\sim 1/M_\pi$
are not affected at two (and higher) loop accuracy. Thus, the Born
contribution gets sizeably reduced at this order. It is also
interesting to note that the formally leading term $\sim \mu^0$ is
very small. Since the proton
pole contributions have the same sign as the ones from the neutron and
are larger in magnitude (related largely to the fact that the magnetic
moment is larger), one can speculate that these terms will sizeably
reduce the positive prediction, cf. Eq.(\ref{gammapn}). Such a
calculation should be done but goes beyond the scope of this article.
Finally, we have considered the moment  $I_A^n (Q^2)$. Although we find some
sizeable fifth order corrections from the Born terms, these are
counterbalanced to a large extent by vector meson contributions,
so that the ensuing band for the theoretical prediction is not very
far from the fourth order one as given in Fig.~\ref{fig:IAndat}

\newpage
\appendix
\def\theequation{\Alph{section}.\arabic{equation}}
\setcounter{equation}{0}
\section{Chiral contributions to the structure functions}
\label{app:form}
Here, we give the explicit expressions for the reduced invariant 
functions $A(s,Q^2)$ and $B(s,Q^2)$ as defined in Eq.(\ref{AB}),
for the third and fourth order contributions. Throughout, the
contribution from the elastic intermediate state is subtracted.
For notational simplicity, we do not display the bar as done for
the structure functions $S_{1,2}^{p,n} (\nu,Q^2)$ in the main text,
cf. Eq.~(\ref{red}). Note that the $B(s,Q^2)$ as given below correspond to the
quantity 
$$\nu \bar {\underline S}_2(\nu,Q^2)=\nu \bar S_2(\nu,Q^2) -
\nu\bar S_2(\nu,Q^2)\big|_{\nu=0}\, ,
$$
that is they allow to calculate
the moments of  $\bar{S}_2$ for $i \ge 1$. However, from the ${\cal O}(q^4)$
graphs which have an elastic piece, i.e 4(a)-4(j), one gets an additional 
inelastic contribution. Indeed, one has:
\beq
B(s,Q^2)-B(2m^2-2Q^2-s,Q^2) = e^2 \frac{\nu}{\nu^2-\nu_B^2} {\tilde B}(Q^2)
+ \ldots ~,
\eeq
where $\nu_B=Q^2/2m$ and the ellipsis denotes the inelastic contribution given
below.
Using the fact that the pole contribution to $\nu S_2$ reads
\beqa
 \nu S_2(\nu,Q^2)^{\rm pole} =\frac{e^2}{2}  \frac{\nu_B^2}{\nu^2-\nu_B^2}
F_2(Q^2)(F_1(Q^2)+F_2(Q^2))
\eeqa
one gets:
\beqa
S_2^{(-1)}(0,Q^2)={\tilde B}(Q^2)=\frac{1}{2}F_2(Q^2)(F_1(Q^2)+F_2(Q^2))
\eeqa
It is straightforward to check that this relation holds.

\medskip\noindent
Consider first the third order diagrams given in Fig.~\ref{fig:diag3} and
their crossed partners. We have (whenever appropriate, we add together
the contributions from various diagrams):

\medskip
\noindent \underline{3a}
\beqa
 A(s,Q^2) &=& {m e^2 g_A^2  \over 2F^2} \, \Bigl[ J_0(s) + {1 \over 2m^2}
 (s+m^2+Q^2) J_1 (s) \Bigr]~, 
 \nonumber \\
 B(s,Q^2) &=& 0~. 
\eeqa
\noindent \underline{3b+3c}
\beqa
 A(s,Q^2) &=& {m e^2 g_A^2 \over F^2} \, \Bigl[ J_0(s) - {1 \over 2m^2}
 (s+m^2+Q^2) J_1 (s) - 2  Q^2 \Gamma_1 (s,Q^2) \nonumber \\
 && +  (s-m^2+Q^2) \Gamma_2 (s,Q^2) + 4 \Gamma_3 (s,Q^2) \Bigr]
 \, \frac{1}{2} (1-\tau_3)~, 
 \nonumber \\
 B(s,Q^2) &=& {2m e^2 g_A^2 \over F^2}  \, \Bigl[ \Gamma_1 (s,Q^2) -
 \Gamma_4 (s, Q^2) \Bigr] \, \frac{1}{2} (1-\tau_3)~. 
\eeqa
\noindent \underline{3d+3e}
\beqa
 A(s,Q^2) &=& {4 m e^2 g_A^2 \over F^2} \, \gamma_3 (s,Q^2)~,
 \nonumber \\
 B(s,Q^2) &=& -{m e^2 g_A^2 \over F^2} \,  \Bigl[ 2\gamma_4 (s,Q^2) +
 \gamma_1 (s, Q^2) \Bigr]~.
\eeqa
\noindent \underline{3f}
\beqa
 A(s,Q^2) &=& {4 m e^2 g_A^2 \over F^2} \, \Bigl[-\gamma_3 (s,Q^2)
 +{1\over 2} (s-m^2+Q^2) \Omega_3 (s,Q^2) \Bigr]\, \frac{1}{2} (1-\tau_3)~,
 \nonumber \\
 B(s,Q^2) &=& {m e^2 g_A^2 \over F^2} \,  \Bigl[ 2\gamma_4 (s,Q^2) +
 \gamma_1 (s, Q^2) + 2m^2 \Omega_5 (s,Q^2) \Bigr]\, \frac{1}{2} (1-\tau_3)~.
\eeqa
\noindent \underline{3g}
\beqa
 A(s,Q^2) &=& {m e^2 g_A^2 \over F^2} \, \Bigl[-\frac{3}{2}  J_0(s)
 +{1\over 4m^2} (s+m^2+Q^2) J_1 (s) + (s-m^2+3Q^2) \Gamma_1(s,Q^2) 
\nonumber \\
&& + (3m^2-Q^2-s) \Gamma_2 (s,Q^2) - 4  \Gamma_3 (s,Q^2)
+ (6-d)  (s-m^2+Q^2) G_3 (s,Q^2)
\nonumber \\
&& -  Q^2 (s-m^2+Q^2) G_4 (s,Q^2) - 4m^2 Q^2  G_5 (s,Q^2)
+ m^2 (s-m^2+Q^2)  G_6 (s,Q^2)  \Bigr]\, \frac{1}{4} (3-\tau_3)~,
\nonumber \\
 B(s,Q^2) &=& -{2 m e^2 g_A^2 \over F^2} \,  \Bigl[ \Gamma_1 (s,Q^2) -
 \Gamma_4 (s, Q^2)  \Bigr]\, \frac{1}{4} (3-\tau_3)~.
\eeqa
\noindent \underline{3h}
\beqa
 A(s,Q^2) &=& {m e^2 g_A^2 \over F^2} \, \Bigl[-\frac{3}{4}  J_0(s)
 +{3\over 8m^2} (s+m^2+Q^2) J_1 (s) - \frac{3 Q^2}{2(s-m^2)} \Bigl(
\underline J_0(s) - 2\underline J_1(s) \Bigr) 
\nonumber\\
&& - \frac{3 Q^2 m^2}{(s-m^2)^2}  \Bigl( \underline {\underline J}_0(s)
- \underline {\underline J}_1(s) \Bigr) \Bigr] \, \frac{1}{2} (1+\tau_3)~,
\eeqa
\noindent \underline{3i+3j}
\beqa
 A(s,Q^2) &=& {m e^2 g_A^2 \over F^2} \, \Bigl[\frac{1}{2m^2} (m^2 -Q^2-s) J_1(s)
 - \frac{Q^2}{s-m^2} \Bigl( \underline J_1(s) - \underline J_0(s) \Bigr) 
\Bigr] \, \frac{1}{2} (1+\tau_3)~,
\eeqa
\noindent \underline{3k+3l}
\beqa
 A(s,Q^2) &=& {m e^2 g_A^2 \over F^2} \, \Bigl(\frac{4 Q^2}{s-m^2}\Bigr) \,
\underline \gamma_3 (s,Q^2) \, \frac{1}{2} (1+\tau_3)~,
\eeqa
\noindent \underline{3m+3n}
\beqa
 A(s,Q^2) &=& {m e^2 g_A^2 \over F^2} \, \Bigl[\frac{1}{4m^2} (m^2 -Q^2-s) J_1(s)
 -Q^2  \Gamma_1 (s, Q^2) + \frac{1}{2} (s-m^2 +Q^2) \Gamma_2 (s, Q^2)
 + Q^2  \Gamma_5 (s, Q^2)
\nonumber\\
&& \qquad\qquad - \frac{Q^2}{s-m^2} \Bigl( \frac{1}{2}\underline J_1(s) 
- \frac{1}{2} \underline J_0(s) - (d-2) \underline \Gamma_3 (s,Q^2) + Q^2
\underline \Gamma_4 (s,Q^2) 
\nonumber\\
&& \qquad \qquad \qquad
- Q^2 \underline \Gamma_5 (s,Q^2) - m^2 \underline \Gamma_6 (s,Q^2) \Bigr) 
\Bigr] \, \frac{1}{2} (1+\tau_3)~.
\eeqa
\noindent \underline{3h+3i+3j+3k+3l+3m+3n}
\beqa
 B(s,Q^2) &=& {m e^2 g_A^2 \over F^2} \, \Bigl[ \gamma_1 (s,Q^2)
 + 2\gamma_4 (s,Q^2) + \frac{2m^2}{s-m^2} \Bigl( \underline \gamma_2 (s,Q^2)
 + 2 \underline \gamma_5 (s,Q^2) \Bigr)\Bigr]\, \frac{1}{2} (1+\tau_3)~,
\eeqa
with
\beqa
 \underline F(s,Q^2) &=&  F(s,Q^2) - F(m^2, Q^2)~, \nonumber \\
\underline {\underline F}(s,Q^2) &=&  \underline F(s,Q^2) - (s-m^2) 
\underline F^\prime (m^2, Q^2)~,
\eeqa
where the prime denotes differentiation with respect to $s$ with $s=
 (p+q)^2 = 2m \nu + m^2 - Q^2$ the standard Mandelstam variable and 
$\nu$ is the energy transfer. 
The various loop functions are generalizations of the ones defined for $Q^2=0$ 
in Ref.~\cite{BKMpol} to virtual photons with $Q^2>0$ using 
\beq
h_\gamma(x,y;s,Q^2) = M_\pi^2(1-y) +M^2y^2+(s-M^2)xy(y-1) +Q^2x(1-x)(1-y)^2
\eeq 
in their Feynman parameter representation. Note that the result for
$A(s,Q^2)$ agrees (up to an overall factor of $m$) with the relativistic 
one given in Appendix~A of Ref.~\cite{EKPW} if one uses the relations
\beqa
(s-m^2+Q^2) \Omega_3 (s,Q^2) &=& \gamma_3 (s, Q^2) +  \Gamma_3 (s, Q^2)
- (s \to 2m^2 - 2Q^2 -s)~, \nonumber \\
2 G_3 (s,Q^2) &=& \Gamma_4 (s, Q^2) -  \Gamma_2 (s, Q^2)~, 
\eeqa
that hold in dimensionally regularized but {\em not} in infrared regularized 
baryon CHPT.  

\medskip\noindent
Consider now  the fourth order diagrams given in Fig.~\ref{fig:diag4} and
their crossed partners. After properly taking into account  the renormalization
of the anomalous magnetic moment \cite{KM}, we find (whenever appropriate, we 
add together the contributions from various diagrams. In particular, we have 
not shown the diagrams which arise from a trivial reordering of the two vertices,
like e.g. for the contribution from graphs 4e)+4f). Such contributions are included 
in the results given.):

\medskip
\noindent \underline{4a+4b+4c+4d}
\beqa
 A^n(s,Q^2) &=& {e^2 g_A^2 \over 4m F^2} \, {1 \over 2} \, (\kappa_s
 -\kappa_v)  \Bigl\{ -2(s-m^2+Q^2) \Bigl[ (s-m^2) (\Gamma_1 (s,Q^2)
 - \Gamma_5 (s,Q^2)) + (2-d) \Gamma_3 (s,Q^2) 
\nonumber \\
&& + Q^2 (\Gamma_4 (s,Q^2)
 - \Gamma_5 (s,Q^2)) - m^2 \Gamma_6 (s,Q^2) + 2\gamma_3 (s,Q^2) \Bigr]
 -4 m^2 Q^2 (2 \Gamma_1 (s,Q^2)  -\Gamma_2 (s,Q^2) 
\nonumber \\
&& - 2\Gamma_5 (s,Q^2) ) - \frac{8Q^2 m^2}{s-m^2} \Bigl[ (2-d) \underline
 \Gamma_3 (s,Q^2) + Q^2 ( \underline \Gamma_4 (s,Q^2) - \underline
 \Gamma_5 (s,Q^2) )
\nonumber \\
&& - m^2  \underline \Gamma_6 (s,Q^2) +2 \gamma_3(s,Q^2) \Bigr] \Bigl\} 
\nonumber \\
 B^n(s,Q^2) &=& {e^2 g_A^2 \over 4m F^2} \, {1 \over 2} \, (\kappa_s
 -\kappa_v)  \Bigl\{ 2 m^2 \Bigl[ 6 \Gamma_1 (s,Q^2) -\Gamma_2 (s,Q^2)
 - 4\Gamma_4 (s,Q^2) - 4 \Gamma_5 (s,Q^2) - 2\gamma_1 (s,Q^2)
\nonumber \\
&&  - \gamma_2 (s,Q^2) - 4\gamma_4(s,Q^2)  - 2\gamma_5 (s,Q^2) \Bigr]
    + 2J_{21}^{\pi\pi} (Q^2) - 3J_1^{\pi\pi} (Q^2) + J_{\pi\pi} (Q^2)
\nonumber \\
&& + \frac{2 m^2}{s-m^2} \Bigl[2 (2-d) \underline
 \Gamma_3 (s,Q^2) + 2 Q^2  \underline \Gamma_4 (s,Q^2) - 2(4m^2 +Q^2)  \underline
 \Gamma_5 (s,Q^2)  -2 m^2 \underline \Gamma_6 (s,Q^2)
\nonumber \\
&& + 4  \underline \gamma_3 (s,Q^2) - 4m^2 ( \underline \gamma_2
 (s,Q^2) + 2 \underline \gamma_5 (s,Q^2) )  \Bigr] \Bigl\}~, 
\nonumber \\
 A^p(s,Q^2) &=& {e^2 g_A^2 \over 4m F^2} \, {1 \over 2} \, (\kappa_s
 +\kappa_v)  \Bigl\{ -(s-m^2+Q^2) \Bigl[ -\frac{3}{2} (J^{\pi N} (s)
 - J_1^{\pi N} (s) ) + (s-m^2) \Gamma_1 (s,Q^2) 
\nonumber \\
&&+ (2-d) \Gamma_3 (s,Q^2)  + Q^2 \Gamma_4 (s,Q^2)
 + (m^2-Q^2-s)  \Gamma_5 (s,Q^2) - m^2 \Gamma_6 (s,Q^2) -4\gamma_3
 (s,Q^2) \Bigr]
\nonumber \\
&& -3  Q^2 J_1^{\pi N} (s) - 2 m^2 Q^2( 2 \Gamma_1 (s,Q^2) - \Gamma_2
 (s,Q^2)  - 2\Gamma_5 (s,Q^2) ) 
\nonumber \\
&& - \frac{4Q^2 m^2}{s-m^2} \Bigl[ -\frac{3}{2} (\underline J^{\pi N}
 (s) - \underline J_1^{\pi N} (s) )
 + (2-d) \underline \Gamma_3
 (s,Q^2) + Q^2 ( \underline \Gamma_4 (s,Q^2) - \underline
 \Gamma_5 (s,Q^2) )
\nonumber \\
&& \qquad \qquad \qquad 
- m^2  \underline \Gamma_6 (s,Q^2) -4 \gamma_3(s,Q^2) \Bigr] \Bigl\} 
\nonumber \\
 B^p(s,Q^2) &=& - {e^2 g_A^2 \over 4m F^2} \, {1 \over 2} \, (\kappa_s
 +\kappa_v)  \Bigl\{ 2 m^2 \Bigl[ 3 \Gamma_1 (s,Q^2) - \frac{1}{2}\Gamma_2 (s,Q^2)
 - 2\Gamma_4 (s,Q^2)) - 2 \Gamma_5 (s,Q^2) + 2\gamma_1 (s,Q^2)
\nonumber \\
&&  + 4\gamma_4(s,Q^2)  \Bigr] +\frac{3}{2}J_1^{\pi N}(s)
    - 2J_{21}^{\pi\pi} (Q^2) + 3J_1^{\pi\pi} (Q^2) - J_{\pi\pi} (Q^2)
    + \frac{2 m^2}{s-m^2} \Bigl[ -\frac{3}{2} ( \underline J^{\pi N} (s)
    - \underline J_1^{\pi N} (s) )
\nonumber \\
&& +  (2-d) \underline \Gamma_3 (s,Q^2) - Q^2 (  \underline \Gamma_5
 (s,Q^2) - \underline \Gamma_4 (s,Q^2) ) - m^2 (\underline \Gamma_6
 (s,Q^2) + 4\underline \Gamma_5 (s,Q^2) )
\nonumber \\
&& \qquad \qquad  +4m^2(  \underline \gamma_2 (s,Q^2) +2 \underline \gamma_5 (s,Q^2))
 - 4 \underline \gamma_3 (s,Q^2)   \Bigr] \Bigl\}~. 
\eeqa
\underline{4g}
\beqa
A^p (s,Q^2) &=&  {e^2 g_A^2 \over 4m F^2} \, \Bigl(-{3 \over 4}\Bigr) \, (\kappa_s
 +\kappa_v)  \Bigl[ -\frac{16 m^4 Q^2}{(s-m^2)^2} \Bigl( \underline
 {\underline J}_1 (s) - \underline {\underline J}_0 (s) \Bigr) 
 - \frac{4m^2 Q^2}{s-m^2} \Bigl( \underline  {\underline J}_1 (s) 
 - \underline {\underline J}_0 (s) + 4 \underline J_1 (s) - 2
 \underline J_0 \Bigr)
\nonumber \\
&& -4m^2 ( \underline  {\underline J}_1 (s) - \underline 
   {\underline J}_0 (s) ) - (s-m^2+Q^2) (3 \underline J_1 (s)
   - \underline J_0 (s) )  - 2Q^2 J_1 (s) \Bigr] ~,
\nonumber \\ 
B^p (s,Q^2) &=&  {e^2 g_A^2 \over 4m F^2} \, {3 \over 4} \, (\kappa_s
 +\kappa_v)  \Bigl[ \frac{8 m^4}{(s-m^2)^2} \Bigl( \underline
 {\underline J}_1 (s) - \underline {\underline J}_0 (s) \Bigr)
 + \frac{4m^2}{s-m^2} \Bigl( 2\underline J_1 (s) -
 \underline J_0 (s) \Bigr) + J_1 (s) \Bigr]~.  
\eeqa
\underline{4j}
\beqa
A (s,Q^2) &=&  {e^2 g_A^2 \over 4m F^2} \, {1 \over 4} \,(1+\tau_3)
 \, (3\kappa_s- \kappa_v) \Bigl\{ -(s-m^2+Q^2) \Bigl[ J_1 (s) - J_0 (s)
 +2(s-m^2+Q^2) \Gamma_2 (s,Q^2) 
\nonumber \\
&& \qquad - 2M_\pi^2 \Gamma_0 (s,Q^2) + 4\Gamma_3 (s,Q^2) \Bigr] +2Q^2 \Bigl( J_1 (s)
-4m^2 \Gamma_1 (s,Q^2) \Bigr) 
\nonumber \\
&& \qquad\qquad- {8 Q^2 m^2 \over s-m^2} \Bigl( 
2\underline \Gamma_3 (s,Q^2) - M_\pi^2 \underline \Gamma_0 (s,Q^2) 
\Bigr) \Bigr\} 
\nonumber \\
B  (s,Q^2) &=&  {e^2 g_A^2 \over 8 m F^2} \,{1 \over 4} \,(1+\tau_3)
 \, (3\kappa_s- \kappa_v) \Bigl\{ J_1 (s) - 2m^2 (\Gamma_2 (s,Q^2)
+ 2\Gamma_5 (s,Q^2) )  
\nonumber \\
&& + {4m^2 \over s-m^2} \Bigl[ \frac{1}{4} \Bigl( J_1 (s) - J_0 (s))
+ M_\pi^2 \Gamma_0 (s,Q^2) - 4\Gamma_3 (s,Q^2) + 2Q^2 
(\underline \Gamma_4 (s,Q^2) - \underline \Gamma_5 (s,Q^2) ) \Bigr]
 \Bigr\}~.
\nonumber \\
&&
\eeqa
\underline{4k+4l}
\beqa
A (s,Q^2) &=&  {e^2 g_A^2 \over 4m F^2} \, {1 \over 2} 
 \, (-\kappa_v + \kappa_s \tau_3) \Bigl\{ (s-m^2+Q^2) \Bigl[ J_0 (s) + J_1 (s)
 -2 M_\pi^2 \Gamma_0 (s,Q^2) + 2Q^2 \Gamma_1 (s,Q^2) 
\nonumber \\
&& - 4m^2 \Gamma_2 (s,Q^2)  + 8\Gamma_3 (s,Q^2) - 4Q^2 \Gamma_4 (s, Q^2) 
 + 2(s-m^2+Q^2) \Gamma_5 (s,Q^2) +4 m^2 \Gamma_6 (s,Q^2) \Bigr]
\nonumber \\
&& - 2Q^2 \Bigl( J_1 (s) -2 m^2 \Bigl( 2\Gamma_1 (s,Q^2) + \Gamma_2
 (s,Q^2) -2 \Gamma_5 (s,Q^2) \Bigr) \Bigr) \Bigr\} 
\nonumber \\
B  (s,Q^2) &=& - {e^2 g_A^2 \over 4 m F^2} \,{1 \over 2} \,
 \, (-\kappa_v + \kappa_s \tau_3) \Bigl[ J_1 (s) - 2m^2 \Bigl(2\Gamma_1 (s,Q^2)
+ \Gamma_2 (s,Q^2) - 2\Gamma_5 (s,Q^2) \Bigl) \Bigr]~.
\eeqa
\underline{4m}
\beqa
A (s,Q^2) &=&  {e^2 g_A^2 \over 4m F^2} \, {1 \over 2} 
 \, (-\kappa_v + \kappa_s \tau_3) \Bigl[4 (s-m^2+Q^2)  \gamma_3 (s,Q^2)
 -16 m^2 \Gamma_3 (s,Q^2) \Bigr]~,
\nonumber \\
B  (s,Q^2) &=&  {e^2 g_A^2 \over 4 m F^2} \,{1 \over 2} \,
 \, (-\kappa_v + \kappa_s \tau_3) 4m^2 \, \Bigl[\Gamma_1 (s,Q^2)
 - 2\Gamma_4 (s,Q^2) - {1\over 2} \gamma_2 (s,Q^2) -\gamma_5 (s,Q^2) 
 + M_\pi^2 \Omega_1 (s,Q^2) \Bigr]~. \nonumber \\ &&
\eeqa
\underline{4n+4o}
\beqa
A (s,Q^2) &=&  {e^2 g_A^2 \over 4m F^2} \, {1 \over 4}\, (3-\tau_3) 
 \, (\kappa_s + \kappa_v) \Bigl\{ (s-m^2+Q^2) \Bigl[ {1\over 2} 
 \Bigl( J_0 (s) + J_1 (s) \Bigr) - (s-m^2-Q^2)  \Gamma_1 (s,Q^2)
\nonumber \\
&& -2 m^2 \Gamma_2 (s,Q^2) + 2\Gamma_3 (s,Q^2) -2Q^2 \Gamma_4 (s,Q^2)
   +(s-m^2+Q^2) \Gamma_5 (s,Q^2) + 2m^2 \Gamma_6 (s,Q^2)
\nonumber\\
&&   + 8m^2 G_3 (s,Q^2) \Bigr] - Q^2 \Bigl( J_1(s) - 4m^2 ( \Gamma_2
 (s,Q^2) - \Gamma_5 (s,Q^2) - M_\pi^2 G_0 (s,Q^2) + 2G_3 (s,Q^2) ) \Bigr)
\nonumber\\
&& \qquad\qquad\qquad - 8m^2 ( \Gamma_3 (s,Q^2) - \Gamma_3 (M^2,Q^2))
 \Bigr\}~,
\nonumber\\
 B  (s,Q^2) &=&  {e^2 g_A^2 \over 4 m F^2} \,{1 \over 2} \,
 (3-\tau_3)  \, (\kappa_s + \kappa_v) \, \Bigl\{ -\frac{1}{4} J_1 (s)
 + m^2 \Bigl[ \Gamma_2 (s,Q^2) - 2 \Gamma_4 (s,Q^2) 
 - M_\pi^2 ( G_0 (s,Q^2) 
\nonumber \\
&& + 2G_1 (s,Q^2))  + 4 G_3 (s,Q^2) - (m^2 -Q^2
 -s) ( 2G_4 (s,Q^2) - G_5 (s,Q^2)) \Bigr] \Bigr\}~.
\eeqa

\pagebreak

\pagebreak

\section*{Figures}

\vspace{2.5cm}

\newpage


\newpage

$\,$

\bigskip
 
\end{document}